\documentclass[12pt]{article}
\newcommand{\myappendix}{\setcounter{equation}{0}\appendix}
\usepackage{epsfig}
\usepackage{graphicx}
\usepackage{a4}
\usepackage{latexsym}
\usepackage{cite}

\usepackage{pslatex}
\usepackage[latin1]{inputenc}
\usepackage[T1]{fontenc}
\newcommand{\ep}{\mbox{$\varepsilon$}}

\begin{document}
\setlength{\parskip}{0.2cm}
\setlength{\baselineskip}{0.55cm}
\begin{titlepage}
\begin{flushright} 
\hfill {\tt hep-ph/xxxxxxx}
\\
\hfill {\tt HRI-03/2006}
\end{flushright} 
\vspace{5mm} 
\begin{center} 
{\Large \bf 
Higher-Order threshold effects to inclusive processes in QCD}\\
\end{center}

\vspace{10pt} 
\begin{center} 
{\bf 
V. Ravindran
}\\ 
\end{center}
\begin{center} 
{\it 
Harish-Chandra Research Institute, 
 Chhatnag Road, Jhunsi, Allahabad, India.\\
} 
\end{center}
 
\vspace{10pt} 
\begin{center}
{\bf ABSTRACT} 
\end{center} 
We present threshold enhanced QCD corrections to inclusive processes 
such as Deep inelastic scattering, Drell-Yan process 
and Higgs productions through gluon fusion
and bottom quark annihilation processes using the resummed cross sections.  
The resummed cross sections are derived using
renormalisation group invariance and mass factorisation theorem 
that these hard scattering cross sections satisfy and 
Sudakov resummation of QCD amplitudes. 
We show how these higher order threshold QCD corrections
improve the theoretical predictions for 
the Higgs production through gluon fusion at hadron colliders.

\vskip12pt 
\vskip 0.3 cm

\end{titlepage}
Perturbative Quantum Chromodynamics(pQCD) provides a framework  
to successfully compute various observables involving hadrons at
high energies.  Wealth of precise data from hadronic experiments
such as deep inelastic scattering(DIS) experiments at HERA 
complimented with the theoretical advances that lead to enormous
success in computing the relevant observables very precisely has given
us a better understanding of the structure of hadrons and also the
strong interaction dynamics in terms of their constituents 
such as quarks and gluons at wide range of energies.  
With these successes we can now predict
most of the important observables with less theoretical uncertainty
for the physics studies at present collider Tevatron in Fermi-Lab 
as well as at the upcoming Large Hadron Collider(LHC) in CERN
\cite{Dittmar:2005ed}.
 
The Drell-Yan(DY) production of di-leptons 
is one of the important processes at hadronic colliders.
It will serve not only as a luminosity monitor but also 
provide vital information on physics beyond Standard Model(SM). 
The other process which is equally important is Higgs boson production 
at such colliders because it will establish the
Standard Model as well as beyond SM Higgs \cite{Djouadi:2005gi,Djouadi:2005gj}.
In pQCD, the DY production of di-leptons and 
Higgs boson production are known upto next to next to leading order(NNLO) 
level in QCD \cite{Altarelli:1978id,
Dawson:1990zj,Djouadi:1991tk,Spira:1995rr,Matsuura:1987wt,
Matsuura:1988sm,Hamberg:1990np,Harlander:2001is,Catani:2001ic, 
Harlander:2002wh,Anastasiou:2002yz,Ravindran:2003um,Harlander:2003ai}.
Beyond NLO, the Higgs production cross sections are known only
in the large top quark mass limit.  
Apart from these fixed order results, the resummation programs 
for the threshold corrections to 
both DY and Higgs productions have also been very successful
\cite{Sterman:1986aj,Catani:1989ne}(see also \cite{Kodaira:1981nh}).
For next to next to leading logarithmic (NNLL) resummation, see 
\cite{Vogt:2000ci,Catani:2003zt}.
Due to  several important results at three loop level 
that are available in recent times
\cite{Moch:2004pa,Vogt:2004mw,Moch:2005id,Moch:2005tm,Moch:2005ba,Blumlein:2004xt},
the resummation upto $N^3LL$ has also become reality \cite{Moch:2005ky,
Laenen:2005uz,Idilbi:2005ni,Ravindran:2005vv}. 
These results in fixed order as well as 
resummed calculations unravel the 
interesting structures in the perturbative results (for example: 
\cite{Blumlein:2000wh,Blumlein:2005im,Dokshitzer:2005bf,Ravindran:2005vv}).

In the QCD improved parton model, the infra-red safe observables,
say, hadronic cross sections can be expandable in terms of 
perturbatively calculable partonic cross sections 
appropriately convoluted with non-perturbative operator matrix elements
known as parton distribution functions(PDF). 
This is possible due to the factorisation property that certain hard scattering
cross sections satisfy.  The partonic cross sections are calculable
in powers of the strong coupling constant $g_s$ within the framework of
perturbative QCD because the coupling constant becomes small at high energies.
On the other hand we only know the perturbative scale evolution of
the parton distribution functions which otherwise are known/extracted 
from experiments.

The fixed order QCD predictions have limitations in applicability
due to the presence of various logarithms of kinematical origins.  These
logarithms become large in some kinematical regions which otherwise can be
probed by the experiments.  In such regions, the applicability of
fixed order perturbative results becomes questionable due to the missing
higher order corrections that are hard to compute with the present day
techniques.  The alternate approach to probe these regions is 
to resum these logarithms in a closed form.
Such an approach of resuming a class of large logarithms supplemented with 
fixed order results can almost cover the entire kinematic region of
the phase space.  
In addition, these threshold corrections 
are further enhanced when the the flux
of the incoming partons become large in those regions.  In the case
of Higgs production through gluon fusion, the gluon flux at small
partonic energies becomes large improving the role of
threshold corrections.    
Here,we consider the inclusive cross 
sections of hadronic cross reactions 
such as deep inelastic scattering, DY, Higgs production 
through gluon fusion and bottom quark
annihilation and study the effects of soft gluons that originate in the
threshold region of the phase space.
In these processes, large logarithms are generated when the gluons
that are emitted from the incoming/outgoing partons become soft.   

In \cite{Ravindran:2005vv}, we extracted the
soft distribution functions of Drell-Yan and Higgs production
cross sections in perturbative QCD and showed
that they are maximally non-abelien.
That is, we found that the soft distribution function of
Higgs production can be got entirely
from the DY process by a simple multiplication of the colour
factor $C_A/C_F$.  In this article we extend the
applicability of this method to the other important process, namely
Higgs production through bottom quark annihilation.
Using the soft distribution functions extracted
from DY, and the form factor of the Yukawa coupling of Higgs to
bottom quarks, we can now predict soft plus virtual part of
the Higgs production through bottom quark annihilation beyond NNLO
with the same accuracy that DY process and the gluon fusion to
Higgs are known \cite{Moch:2005ky,Ravindran:2005vv}.
This generalises our earlier approach
to include any infrared safe inclusive cross section. 
In \cite{Ravindran:2005vv} we  
determined the threshold exponents $D_i^I$ upto three
loop level for DY and Higgs productions using our resummed
soft distribution functions.  In this present work we
provide all order proof which establishes the relation 
between soft distribution functions
and the threshold resummation exponents and demonstrate the usefulness of this
approach to derive higher order threshold enhanced corrections 
for any infrared safe inclusive cross sections.
The results on DIS in this article provide a consistency check
on various approaches (see \cite{Moch:2005ky,Ravindran:2005vv}) 
in the literature to study the soft part of
the cross sections because some of the constants that go into our study
were extracted from the deep inelastic scattering coefficient functions
\cite{Moch:2005tm}.
In the following, we systematically formulate a framework
to resum the dominant soft gluon contributions in $z$ space 
to the inclusive cross sections.
Here the variable $z$ is the appropriate
scaling variable that enters the cross sections.  To achieve this, 
we use renormalisation group(RG) invariance, mass factorisation and 
Sudakov resummation of QCD amplitudes as the guiding principles.    
Using the resummed results in $z$ space, we predict soft plus 
virtual part of the dominant partonic cross sections beyond 
$N^2LO$.  The soft plus virtual corrections are also called
threshold corrections.  At the end, we show how these results affect the total
cross section and improve the scale uncertainty for the Higgs
production through gluon fusion.  

Since we are only interested in the effect of soft gluons, the
infra-red safe observable can be obtained by adding soft part of
the cross sections with the virtual contributions and performing 
mass factorisation using appropriate counter terms.  
We call this infra-red safe combination a "soft plus virtual"($sv$) 
part of the cross section.      
The soft plus virtual part of the cross section
($\Delta^{sv}_{~I,P}(z,q^2,\mu_R^2,\mu_F^2)$)
after mass factorisation is found to be
\begin{eqnarray}
\Delta^{sv}_{~I,P}(z,q^2,\mu_R^2,\mu_F^2)={\cal C} \exp
\Bigg({\Psi^I_P(z,q^2,\mu_R^2,\mu_F^2,\ep)}\Bigg)\Bigg|_{\ep=0}
\label{master}
\end{eqnarray}
where $\Psi^I_P(z,q^2,\mu_R^2,\mu_F^2,\ep)$ is a finite distribution.
The subscript $P=S$ for Drell-Yan(DY) and Higgs productions and 
$P=SJ$ for deep inelastic scattering.  The symbol $S$ stands for "soft"
and $SJ$ stands for "soft plus jet".   For DY and DIS, $I=q$(quark/anti-quark) 
and for Higgs production through gluon fusion, $I=g$(gluon)
and for bottom quark annihilation to Higgs boson, $I=b$(bottom quark).  
Here $\Psi^I_P(z,q^2,\mu_R^2,\mu_F^2,\ep)$  
is computed in $4+\ep$ dimensions.
\begin{eqnarray}
\Psi^I_P(z,q^2,\mu_R^2,\mu_F^2,\ep)&=&
\Bigg(
\ln \Big(Z^I(\hat a_s,\mu_R^2,\mu^2,\ep)\Big)^2
+\ln \big|\hat F^I(\hat a_s,Q^2,\mu^2,\ep)\big|^2
\Bigg)
\delta(1-z)
\nonumber\\[2ex]
&&+2~ \Phi^{~I}_P(\hat a_s,q^2,\mu^2,z,\ep)
-2~m~ {\cal C}\ln \Gamma_{II}(\hat a_s,\mu^2,\mu_F^2,z,\ep),
\quad \quad I=q,g,b 
\label{DYH}
\end{eqnarray}
In the above $m=1$ for DY and Higgs productions and $m=1/2$ for DIS. 
The symbol "${\cal C}$" 
means convolution.  For example, ${\cal C}$ acting on an exponential
of a function $f(z)$ has the following expansion:
\begin{eqnarray}
{\cal C}e^{\displaystyle f(z) }= \delta(1-z)  + {1 \over 1!} f(z)
 +{1 \over 2!} f(z) \otimes f(z) + {1 \over 3!} f(z) \otimes f(z) \otimes f(z)
+ \cdot \cdot \cdot
\end{eqnarray}
In the rest of the paper, the function $f(z)$ 
is a distribution of the kind $\delta(1-z)$ and ${\cal D}_i$, where
\begin{eqnarray}
{\cal D}_i=\Bigg[{\ln^i(1-z) \over (1-z)}\Bigg]_+
\quad \quad \quad i=0,1,\cdot\cdot\cdot
\end{eqnarray}
and the symbol $\otimes$ means the Mellin convolution.
Since we are only interested in the soft plus virtual part of the
cross sections, we drop all the regular functions that result from
various convolutions.
$\hat F^I(\hat a_s,Q^2,\mu^2)$ are the form factors
that enter in the Drell-Yan($I=q$) production, Higgs($I=g,b$) production 
and DIS($I=q/\overline q$) cross sections.  
In the form factors, $Q^2=-q^2$.
For the DY, $q^2=M^2_{l^+l^-}$ is the invariant mass of the final state
di-lepton pair and for the Higgs production,
$q^2=m_H^2$, where $m_H$ is the mass of the Higgs boson.  
For DIS, $q^2=q_{\gamma^*/Z^*}^2$ is the virtuality of the photon or $Z$.
The variable $z$ in DIS is Bj\"orken scaling variable.
In DY and Higgs production, $z$ is the ratio of $q^2$ over $\hat s$,
where $\hat s$ is the center of mass of the partonic system.
The functions $\Phi^{~I}_P(\hat a_s,q^2,\mu^2,z)$ are
called the soft distribution functions.  
The unrenormalised(bare) strong coupling constant
$\hat a_s$ is defined as
\begin{eqnarray}
\hat a_s={\hat g^2_s \over 16 \pi^2}
\end{eqnarray}
where $\hat g_s$ is the strong coupling constant which is dimensionless in
$n=4+\ep$, with $n$ being the number of space time dimensions.
The scale $\mu$ comes from
the dimensional regularisation in order to make the bare coupling constant $\hat g_s$
dimensionless in $n$ dimensions.

The bare coupling constant $\hat a_s$ is related to renormalised one by
the following relation:
\begin{eqnarray}
S_{\ep} \hat a_s = Z(\mu_R^2) a_s(\mu_R^2) \left(\mu^2 \over \mu_R^2\right)^{\ep \over 2}
\label{renas}
\end{eqnarray}
The scale $\mu_R$ is the
renormalisation scale at which the renormalised strong coupling constant
$a_s(\mu_R^2)$ is defined.  The factorisation scale $\mu_F$ is due 
to mass factorisation.
\begin{eqnarray}
S_{\ep}=exp\left\{{\ep \over 2} [\gamma_E-\ln 4\pi]\right\}
\end{eqnarray}
is the spherical factor characteristic of $n$-dimensional regularisation.
\begin{eqnarray}
Z(\mu_R^2)= 1+ a_s(\mu_R^2) {2 \beta_0 \over \ep}
           + a_s^2(\mu_R^2) \Bigg({4 \beta_0^2 \over \ep^2 }+
                  {\beta_1 \over \ep} \Bigg)
           + a_s^3(\mu_R^2) \Bigg( {8 \beta_0^3 \over \ep^3}
                   +{14 \beta_0 \beta_1 \over 3 \ep^2}
                   +{2 \beta_2 \over 3 \ep}\Bigg)
\end{eqnarray}
The renormalisation constant $Z(\mu_R^2)$ relates the bare coupling constant
$\hat a_s$ to the renormalised one $a_s(\mu_R^2)$ through
the eqn.(\ref{renas}).

The coefficients $\beta_0$,$\beta_1$ and $\beta_2$ are
\begin{eqnarray}
\beta_0&=&{11 \over 3 } C_A - {4 \over 3 } T_F n_f
\nonumber \\
\beta_1&=&{34 \over 3 } C_A^2-4 T_F n_f C_F -{20 \over 3} T_F n_f C_A
\nonumber \\
\beta_2&=&{2857 \over 54} C_A^3 
          -{1415 \over 27} C_A^2 T_F n_f
          +{158 \over 27} C_A T_F^2 n_f^2
\nonumber\\[2ex]
&&          +{44 \over 9} C_F T_F^2 n_f^2
          -{205 \over 9} C_F C_A T_F n_f
          +2 C_F^2 T_F n_f
\end{eqnarray}
where the color factors for $SU(N)$ QCD are given by
\begin{eqnarray}
C_A=N,\quad \quad \quad C_F={N^2-1 \over 2 N} , \quad \quad \quad
T_F={1 \over 2}
\end{eqnarray}
and $n_f$ is the number of active flavours.  In the case of the Higgs
production, the number of active flavours is five because the
top degree of freedom is integrated out in the large $m_{top}$ limit.

The factors $Z^I(\hat a_s,\mu_R^2,\mu^2,\ep)$ are the overall
operator renormalisation constants.
For the vector current  $Z^q(\hat a_s,\mu_R^2,\mu^2)=1$,
but the gluon operator gets overall renormalisation
\cite{Chetyrkin:1997un} given by
\begin{eqnarray}
Z^g(\hat a_s,\mu_R^2,\mu^2,\ep)&=&
1+\hat a_s \left({\mu_R^2 \over \mu^2}\right)^{\ep \over 2}
  S_{\ep} ~\Bigg[{2 \beta_0 \over \ep}\Bigg]
 +\hat a_s^2 \left({\mu_R^2 \over \mu^2}\right)^{\ep}
S_{\ep}^2 ~\Bigg[{2 \beta_1 \over \ep} \Bigg]
\nonumber\\[2ex]
&&  +\hat a_s^3 \left ({\mu_R^2 \over \mu^2}\right)^{3{\ep \over 2}}
S_{\ep}^3~ \Bigg[
           {1 \over \ep^2}\Big(-2 \beta_0 \beta_1  \Big)
           +{2 \beta_2 \over \ep}\Bigg]
\nonumber\\[2ex]
&& + \hat a_s^4 \left ({\mu_R^2 \over \mu^2}\right)^{2{\ep}}
S_{\ep}^4~ \Bigg[
{1 \over \ep^3} \left({8 \over 3} \beta_0^2 \beta_1\right)
+{1 \over \ep^2} \left(-{16 \over 3} \beta_0 \beta_2\right)
+{ 1 \over \ep} \left(2 \beta_3\right) 
\Bigg]
\end{eqnarray}
and for the bottom quark coupled to Higgs, we have
\begin{eqnarray}
Z^b(\hat a_s,\mu_R^2,\mu^2,\ep)& =&1
    + \hat a_s \left({\mu_R^2 \over \mu^2}\right)^{{\ep \over 2}} S_{\ep}
       \Bigg[ {1 \over \ep}   \Bigg( 2~ \gamma^b_0 \Bigg)\Bigg]
   +\hat a_s^2 \left({\mu_R^2 \over \mu^2}\right)^{{\ep }} S_{\ep}^2
       \Bigg[ {1 \over \ep^2}   \Bigg( 2~ \Big(\gamma^b_0\Big)^2 
           - 2~ \beta_0~ \gamma^b_0 \Bigg)
\nonumber\\[2ex]
&&       + {1 \over \ep}   \Bigg( \gamma^b_1 \Bigg)\Bigg]
    +\hat a_s^3 \left({\mu_R^2 \over \mu^2}\right)^{3{ \ep \over 2}} S_{\ep}^3
       \Bigg[ {1\over \ep^3}   \Bigg( {4\over 3}~ \Big(\gamma^b_0\Big)^3 
        - 4~ \beta_0~ \Big(\gamma^b_0\Big)^2 
          + {8\over 3}~ \beta_0^2~ \gamma^b_0 \Bigg)
\nonumber\\[2ex]
&&       + {1 \over \ep^2}   \Bigg( 2 ~\gamma^b_0 ~\gamma^b_1 
       - {2\over 3}~ \beta_1~ \gamma^b_0 
       - {8\over 3}~ \beta_0 ~\gamma^b_1 \Bigg)
       + {1 \over \ep}   \Bigg( {2\over 3} ~\gamma^b_2 \Bigg)\Bigg]
\nonumber\\[2ex]
&&    +\hat a_s^4 \left({\mu_R^2 \over \mu^2}\right)^{2{ \ep }} S_{\ep}^4
       \Bigg[ {1 \over \ep^4}   \Bigg( {2\over 3}~ \Big(\gamma^b_0\Big)^4 
        - 4~ \beta_0~ \Big(\gamma^b_0\Big)^3 + {22\over 3}~ \beta_0^2~ 
         \Big(\gamma^b_0\Big)^2 - 4~ \beta_0^3 ~\gamma^b_0 \Bigg)
\nonumber\\[2ex]
&&       + {1 \over \ep^3}   \Bigg( 2~ \Big(\gamma^b_0\Big)^2 ~\gamma^b_1 
        - {4\over 3}~ \beta_1~ \Big(\gamma^b_0\Big)^2 - {22\over 3}~ \beta_0~ 
         \gamma^b_0~ \gamma^b_1 + {8\over 3}~ \beta_0 ~\beta_1 ~\gamma^b_0 
        + 6~ \beta_0^2~ \gamma^b_1 \Bigg)
\nonumber\\[2ex]
&&       + {1 \over \ep^2}   \Bigg( {1\over 2} ~\Big(\gamma^b_1\Big)^2 
         + {4\over 3} ~\gamma^b_0 ~\gamma^b_2 
         - {1\over 3}~ \beta_2~ \gamma^b_0
          - \beta_1~ \gamma^b_1 - 3~ \beta_0~ \gamma^b_2 \Bigg)
       + {1 \over \ep}   \Bigg( {1\over 2}~ \gamma^b_3 \Bigg)\Bigg]
\end{eqnarray}
where the anomalous dimensions $\gamma^b_i$ can be obtained from
the quark mass anomalous dimensions \cite{vanRitbergen:1997va}
\begin{eqnarray}
\gamma^b_0&=& 3 C_F
\nonumber\\[2ex]
\gamma^b_1&=& {3 \over 2} C_F^2
           +{97 \over 6} C_F C_A
           -{10 \over 3} C_F T_F n_f
\nonumber\\[2ex]
\gamma^b_2&=&{129 \over 2} C_F^3 
           -{129 \over 4} C_F^2 C_A
           +{11413 \over 108} C_F C_A^2
           +\Big(-46+48 \zeta_3\Big) C_F^2 T_F n_f
\nonumber\\[2ex]
&&           +\left(-{556 \over 27} -48 \zeta_3\right) C_F C_A T_F n_f
           -{140 \over 27} C_F T_F^2 n_f^2
\end{eqnarray}
The bare form factors $\hat F^I(\hat a_s,Q^2,\mu^2,\ep)$
(without overall renormalisation)
of both fermionic and gluonic operators satisfy
the following differential equation that follows from the gauge
as well as
renormalisation group invariances \cite{Sudakov:1954sw,Mueller:1979ih,
Collins:1980ih,Sen:1981sd}.  In dimensional regularisation,
\begin{eqnarray}
Q^2{d \over dQ^2} \ln \hat {F^I}\left(\hat a_s,Q^2,\mu^2,\ep\right)&=&
{1 \over 2 }
\Bigg[K^I\left(\hat a_s,{\mu_R^2 \over \mu^2},\ep\right)
+ G^I\left(\hat a_s,{Q^2 \over \mu_R^2},{\mu_R^2 \over \mu^2},\ep\right)
\Bigg]
\label{sud1}
\end{eqnarray}
where $K^I$ contains all the poles in $\ep$.
On the other hand, $G^I$ collects rest of the terms that are
finite as $\ep$ becomes zero.  Renormalisation group invariance
(RG) of the $\hat F^I(\hat a_s,Q^2,\mu^2,\ep)$ leads 
\begin{eqnarray}
\mu_R^2 {d \over d\mu_R^2}
K^I\Bigg(\hat a_s,{\mu_R^2 \over \mu^2},\ep\Bigg)=-A^I(a_s(\mu_R^2))
\nonumber\\[2ex]
\mu_R^2 {d \over d\mu_R^2}
G^I\Bigg(\hat a_s,{Q^2\over \mu_R^2},
{\mu_R^2 \over \mu^2},\ep\Bigg)=A^I(a_s(\mu_R^2))
\end{eqnarray}
The quantities $A^I$ are the standard cusp anomalous dimensions and they are
expanded in powers of renormalised strong coupling constant $a_s(\mu_R^2)$ as
\begin{eqnarray}
A^I(\mu_R^2)=\sum_{i=1}^\infty a_s^{i}(\mu_R^2)~ A_i^I
\end{eqnarray}
The total derivative is given by
\begin{eqnarray}
\mu_R^2 {d \over d\mu_R^2} = \mu_R^2 {\partial \over \partial \mu_R^2}
+{d a_s(\mu_R^2) \over d\mu_R^2} {\partial \over \partial a_s(\mu_R^2)}
\end{eqnarray}
The RG equation for $G^I$ can also be solved and the solution is found to be
\begin{eqnarray}
G^I\left(\hat a_s,{Q^2 \over \mu_R^2},{\mu_R^2 \over \mu^2},\ep\right)
&=&G^I\left(a_s(Q^2),1,\ep\right)+ \int_{Q^2 \over \mu_R^2}^1
{d\lambda^2 \over \lambda^2} A^I\left(a_s(\lambda^2 \mu_R^2)\right)
\end{eqnarray}

The finite function $G^I(a_s(Q^2),1,\ep)$ can also be expanded
in powers of $a_s(Q^2)$ as
\begin{eqnarray}
G^I(a_s(Q^2),1,\ep)=\sum_{i=1}^\infty a_s^i(Q^2)~ G^{~I}_i(\ep)
\end{eqnarray}
The solution to
the eqn.(\ref{sud1}) can be obtained as a series expansion in
the bare coupling constant in dimensional regularisation.  The formal solution
upto four loop level can be found in \cite{Moch:2005id,Ravindran:2005vv}

The coefficients $G^{~I}_i(\ep)$ can be found for
both $I=q$ and $I=g$ in \cite{Moch:2005tm} to the required accuracy in
$\ep$.  We have extended this to the form factor corresponding to Yukawa
interaction of Higgs boson to the bottom quarks.  We have presented  
the logarithm of this form factor, $\ln \hat F^b$ in the appendix A.  
All these form factors satisfy
\begin{eqnarray}
G^{~I}_1(\ep)&=&
2 \Bigg(B^I_1 - \delta_{~I,g}~ \beta_0-\delta_{~I,b}~ \gamma^b_0\Bigg) + f_1^I
        +\sum_{k=1}^\infty \ep^k  g^{~I,k}_1
\nonumber \\[2ex]
G^{~I}_2(\ep)&=&
2\Bigg(B_2^I-2 ~\delta_{~I,g} ~\beta_1-\delta_{~I,b}~ \gamma^b_1\Bigg) + f_2^I
        -2 \beta_0  g^{~I,1}_1
        +\sum_{k=1}^\infty \ep^k  g^{~I,k}_2
\nonumber \\[2ex]
G^{~I}_3(\ep)&=&
2 \Bigg(B_3^I - 3~\delta_{I,g}~ \beta_2-\delta_{~I,b} ~\gamma^b_2\Bigg) + f_3^I
        -2 \beta_1  g^{~I,1}_1
-2 \beta_0 \Big(g^{~I,1}_2+2 \beta_0 g^{~I,2}_1\Big)
        +\sum_{k=1}^\infty \ep^k g^{~I,k}_3
\nonumber \\[2ex]
G^{~I}_4(\ep)&=&
2 \Bigg(B_4^I - 4~ \delta_{I,g} \beta_3-\delta_{~I,b}~ \gamma^b_3\Bigg) +f_4^I
-2 \beta_2 g^{~I,1}_1
-2 \beta_1 \Big(
g^{~I,1}_2 + 4 \beta_0 g^{~I,2}_1 \Big)
\nonumber\\[2ex]
&&-2 \beta_0 \Big(g^{~I,1}_3+2 \beta_0 g^{~I,2}_2
                         +4 \beta_0^2 g^{~I,3}_1\Big)
        +\sum_{k=1}^\infty \ep^k g^{~I,k}_4
\label{GI}
\end{eqnarray}
Notice that the single poles of the form factors contain  
the combination $$2 \Bigg(B^{~I}_i - \delta_{I,g}~ i~\beta_{i-1} -\delta_{I,b}
\gamma^b_{i-1}\Bigg) + f^{~I}_i$$ at every order in $\hat a_s$.  
The terms proportional $-2(\delta_{I,g}~i~ \beta_{i-1} +\delta_{I,b}
\gamma^b_{i-1})$ come from the large momentum region of the loop integrals
that are giving ultraviolet divergences.  The poles containing them
will go away when the form factors undergo overall operator
UV renormalisation through the renormalisation constants
$Z^{~I}$ which satisfy RG equations
\begin{eqnarray}
\mu_R^2 {d \over d\mu_R^2} \ln Z^g(\hat a_s,\mu_R^2,\mu^2,\ep) &=&
\sum_{i=1}^\infty a^i_s(\mu_R^2)~ \Big(i~\beta_{i-1}\Big)
\nonumber\\
\mu_R^2 {d \over d\mu_R^2}\ln Z^b(\hat a_s,\mu_R^2,\mu^2,\ep)&=&
\sum_{i=1}^\infty a^i_s(\mu_R^2)~ \gamma^b_{i-1}
\end{eqnarray}
where $\ep \rightarrow 0$ is set.  The constants 
$i~\beta_{i-1}$ and $\gamma^b_{i-1}$ are anomalous dimensions
of the renormalised form factors $F^g$ and $F^b$ respectively. 
The poles of the form factors containing terms proportional to
$g^{~I,k}_{i-1}$ multiplied by $\beta_{l}$
are due to coupling constant renormalisation.
It is now straightforward to apply this procedure to 
form factor of many operators in QCD that are of interest.  
After the overall operator renormalisation through $Z^{~I}$ and coupling
constant renormalisation through $Z$, the remaining poles will contain only 
$B^{~I}_i$ and $f^{~I}_i$ in addition to the standard cusp anomalous
dimensions $A^{~I}_i$.  
The constants $B_i^I$ are 
known upto order $a_s^3$ thanks to the recent computation
of three loop anomalous dimensions/splitting functions
\cite {Moch:2004pa,Vogt:2004mw}.  They are found to be
flavour independent, that is $B_i^q=B_i^b$.
The constants $f_i^I$ are analogous to the cusp anomalous dimensions
$A_i^I$ that enter the form factors.  The cusp anomalous dimensions
are flavour independent, $A_i^q=A_i^b$.  It was first noticed in
\cite{Ravindran:2004mb}
that the single pole (in $\ep$) of the logarithm of form factors
upto two loop level ($a_s^2$) can be
predicted due the presence of constants $f_i^I$ because these
$f^{~I}_i$ are found to be maximally non-abelien obeying the relation
\begin{eqnarray}
f_i^q=f_i^b={C_F\over C_A} f_i^g
\end{eqnarray}
similar to $A_i^I$.  
In \cite{Moch:2005tm},
this relation has been found to hold even at the three loop level.  
From the experience upto three loop level, we can now
predict all the poles of the form factor at every order in $\hat a_s$
from these constants $A^I$,$B^I$,$f^I$, their anomalous dimensions
and the finite parts of lower order(in $\hat a_s$) contributions to the form factor.
The last equality in eqn.(\ref{GI}) has been expanded again in terms of
$f^I_4$ and $g^{~I,k}_i$ with $i,k=1,2,3$
so that the single pole can again be predicted.  
One is not sure whether this structure will go through
beyond three loop until an explicit calculation is available.

The collinear singularities that arise due to massless partons are
removed in $\overline {MS}$ scheme using 
the mass factorisation kernel $\Gamma(z,\mu_F^2,\ep)$ as shown
in the eqn.(\ref{DYH}).   We have suppressed the dependence on
$\hat a_s$ and $\mu^2$ in $\Gamma$.
The kernel $\Gamma(z,\mu_F^2,\ep)$ satisfies the 
following renormalisation group equation:  
\begin{eqnarray}
\mu_F^2 {d \over d\mu_F^2}\Gamma(z,\mu_F^2,\ep)={1 \over 2}  P
                         \left(z,\mu_F^2\right)
                        \otimes \Gamma \left(z,\mu_F^2,\ep\right)
\end{eqnarray}
The $P(z,\mu_F^2)$ are well known Altarelli-Parisi splitting
functions(matrix valued) known upto three loop level \cite{Moch:2004pa,Vogt:2004mw}:
\begin{eqnarray}
P(z,\mu_F^2)=
\sum_{i=1}^{\infty}a_s^i(\mu_F^2) P^{(i-1)}(z)
\end{eqnarray}
The diagonal terms of splitting functions
$P^{(i)}(z)$ have the following structure
\begin{eqnarray}
P^{(i)}_{II}(z) = 2\Bigg[ B^I_{i+1} \delta(1-z)
                  + A^I_{i+1} {\cal D}_0\Bigg] + P_{reg,II}^{(i)}(z)
\end{eqnarray}
where $P_{reg,II}^{(i)}$ are regular when the argument takes
the kinematic limit(here $z \rightarrow 1$).
The RG equation of the kernel
can be solved in dimensional regularisation
in powers of strong coupling constant.  Since we are interested
only in the soft plus virtual part of the cross section, only the diagonal parts
of the kernels contribute.  In the $\overline{MS}$ scheme,
the kernel contains only poles in $\ep$.  The kernel can be expanded in powers
of bare coupling $\hat a_s$ as
\begin{eqnarray}
\Gamma(z,\mu_F^2,\ep)=\delta(1-z)+\sum_{i=1}^\infty \hat a_s^i
\left({\mu_F^2 \over \mu^2}\right)^{i {\ep \over 2}}S^i_{\ep}
\Gamma^{(i)}(z,\ep)
\end{eqnarray}
The constants $\Gamma^{(i)}(z,\ep)$ expanded in negative powers of
$\ep$ upto four loop level can be found in \cite{Ravindran:2005vv}.
The $\Gamma_{II}(\hat a_s,\mu_F^2,\mu^2,z,\ep)$ in the eqn.(\ref{DYH})
is the diagonal element of $\Gamma(z,\mu_F^2,\ep)$.

The fact that $\Delta^{sv}_{~I,P}$ are finite in the limit 
$\ep \rightarrow 0$ implies
that the soft distribution functions have pole structure in $\ep$ similar to that
of $\hat F^I$ and $\Gamma_{II}$.  
Hence it is natural to expect that the soft
distribution functions also satisfy Sudakov type
differential equation that
the form factors $\hat F^I$ satisfy(see eqn.(\ref{sud1})):
\begin{eqnarray}
q^2 {d \over dq^2}\Phi^{~I}_P(\hat a_s,q^2,\mu^2,z,\ep) =
{1 \over 2 }
\Bigg[\overline K^{~I}_P\left(\hat a_s,{\mu_R^2 \over \mu^2},z,\ep\right)
+ \overline G^{~I}_P\left(\hat a_s,{q^2 \over \mu_R^2},
{\mu_R^2 \over \mu^2},z,\ep\right)
\Bigg]
\label{sud2}
\end{eqnarray}
where again the constants $\overline K^{~I}_P$ contain all the singular terms and
$\overline G^{~I}_P$s
are finite functions of $\ep$.  
Also, $\Phi^{~I}_P(\hat a_s,q^2,\mu^2,z)$ satisfy
the renormalisation group equation:
\begin{eqnarray}
\mu_R^2 {d \over d\mu_R^2}\Phi^{~I}_P(\hat a_s,q^2,\mu^2,z,\ep)=0
\end{eqnarray}
This renormalisation group invariance leads to
\begin{eqnarray}
\mu_R^2 {d\over d\mu_R^2} \overline K^{~I}_P
\Bigg(\hat a_s, {\mu_R^2 \over \mu^2},z,\ep\Bigg)=
-\overline A^{~I}(a_s(\mu_R^2)) \delta(1-z)
\nonumber\\[2ex]
\mu_R^2 {d \over d\mu_R^2} \overline G^{~I}_P
\Bigg(\hat a_s,{q^2 \over \mu_R^2},{\mu_R^2 \over \mu^2},z,\ep\Bigg)
=\overline A^{~I}(a_s(\mu_R^2)) \delta(1-z)
\end{eqnarray}
If $\Phi^{~I}_P(\hat a_s,q^2,\mu^2,z,\ep)$ have to contain the right poles
to cancel the poles
coming from $\hat F^I$,$Z^I$ and $\Gamma_{II}$ in order to
make $\Delta^{sv}_{~I,P}$ finite, then the constants $\overline A^{~I}$ have to satisfy
\begin{eqnarray}
\overline A^{~I}=-A^I
\end{eqnarray}
Using the above relation, the solution to RG equation for $\overline G^{~I}_P
\left(\hat a_s,{q^2 \over \mu_R^2},{\mu_R^2 \over \mu^2},z,\ep\right)$ is found to be 
\begin{eqnarray}
\overline G^{~I}_P\left(\hat a_s,{q^2 \over \mu_R^2},
{\mu_R^2 \over \mu^2},z,\ep\right)
&=&\overline G^{~I}_P \left(a_s(\mu_R^2),{q^2 \over \mu_R^2},z,\ep\right)
\nonumber\\[2ex]
&=&\overline G^{~I}_P\left(a_s(q^2),1,z,\ep\right)
- \delta(1-z) \int_{{q^2 \over \mu_R^2}}^1
{d\lambda^2 \over \lambda^2} A^I\left(a_s(\lambda^2 \mu_R^2)\right)
\end{eqnarray}
With these solutions, it is straightforward to solve the Sudakov
differential equation:
\begin{eqnarray}
\Phi^{~I}_P(\hat a_s,q^2,\mu^2,z,\ep) &=& \Phi^{~I}_P(\hat a_s,q^2 (1-z)^{2m},\mu^2,\ep)
\nonumber\\[2ex]
&=&\sum_{i=1}^\infty \hat a_s^i \left({q^2 (1-z)^{2m} 
\over \mu^2}\right)^{i {\ep \over 2}} S_{\ep}^i 
\left({i~m~ \ep \over 1-z} \right)\hat \phi_P^{~I,(i)}(\ep)
\end{eqnarray}
where
\begin{eqnarray}
\hat \phi^{~I,(i)}_P(\ep)=
{1 \over i \ep} \Bigg[ \overline K^{~I,(i)}(\ep) 
+ \overline {G}^{~I,(i)}_P(\ep)\Bigg]
\end{eqnarray}
The above solution depends on $m$ so that $\Psi^I_P$ or equivalently
$\Delta^{sv}_{I,P}$ is finite as $\ep \rightarrow 0$(see eqn.(\ref{DYH})).  
The constants $\overline K^{~I,(i)}(\ep)$ are independent of 
$P$ and are determined by expanding
$\overline K^I_P$ in powers of bare coupling constant $\hat a_s$ as
\begin{eqnarray}
\overline K^I_P\left(\hat a_s,{\mu_R^2\over \mu^2},z,\ep\right)
=\delta(1-z) \sum_{i=1}^\infty \hat a_s^i
\left({\mu_R^2 \over \mu^2}\right)^{i {\ep \over 2}}S^i_{\ep}~
\overline K^{~I,(i)}(\ep)
\end{eqnarray}
and solving RG equation for
$\overline K^I_P\left(\hat a_s,{\mu_R^2\over \mu^2},z,\ep\right)$.  We obtain
\begin{eqnarray}
\overline K^{~I,(1)}(\ep)&=& {1 \over \ep} \Bigg(2 A_1^I\Bigg)
\nonumber\\[2ex]
\overline K^{~I,(2)}(\ep)&=& {1 \over \ep^2} \Bigg(-2 \beta_0 A_1^I\Bigg)
                 +{1 \over \ep}\Bigg( A_2^I\Bigg)
\nonumber\\[2ex]
\overline K^{~I,(3)}(\ep)&=& {1 \over \ep^3} \Bigg({8 \over 3} \beta_0^2 A_1^I\Bigg)
                 +{1 \over \ep^2} \Bigg(-{2 \over 3} \beta_1 A_1^I
                        -{8 \over 3} \beta_0 A_2^I \Bigg)
                 +{1 \over \ep} \Bigg({2 \over 3} A_3^I \Bigg)
\nonumber\\[2ex]
\overline K^{~I,(4)}(\ep)&=& {1 \over \ep^4} \Bigg(-4 \beta_0^3 A_1^I\Bigg)
                 +{1 \over \ep^3} \Bigg({8 \over 3} \beta_0 \beta_1 A_1^I
                      + 6 \beta_0^2 A_2^I \Bigg)
\nonumber\\[2ex]
&&                 +{1 \over \ep^2} \Bigg(-{1 \over 3} \beta_2 A_1^I
                       -\beta_1 A_2^I - 3 \beta_0 A_3^I \Bigg)
                   +{1 \over \ep} \Bigg({1 \over 2} A_4^I\Bigg)
\end{eqnarray}
The constants $\overline {G}^{~I,(i)}_P(\ep)$ are related to the finite
function $\overline G^I_P(a_s(q^2),1,z,\ep)$ through the distributions
$\delta(1-z)$ and ${\cal D}_j$.  Defining $\overline {\cal G}_i^I(\ep)$
through 
\begin{eqnarray}
\sum_{i=1}^\infty \hat a_s^i 
\left( {q^2 (1-z)^{2 m} \over \mu^2}\right)^{i{\ep \over 2}} 
S^i_{\ep}
\overline G_P^{~I,(i)}(\ep)
&=&
\sum_{i=1}^\infty a_s^i\left(q^2 (1-z)^{2m}\right) 
\overline {\cal G}^{~I}_{P,i}(\ep)
\label{Gbar1}
\end{eqnarray}
we find
\begin{eqnarray}
\overline { G}^{~I,(1)}_P(\ep)&=&\overline {\cal G}_{P,1}^{~I}(\ep)
\nonumber\\[2ex]
\overline { G}^{~I,(2)}_P(\ep)&=&{1\over \ep} \Bigg(
                  - 2 \beta_0  \overline {\cal G}_{P,1}^{~I}(\ep)\Bigg)
                  +\overline {\cal G}_{P,2}^{~I}(\ep)
\nonumber\\[2ex]
\overline { G}^{~I,(3)}_P(\ep)&=&  {1\over \ep^2} \Bigg(
                    4 \beta_0^2 \overline {\cal G}_{P,1}^{~I}(\ep)\Bigg)
                  +{1\over \ep} \Bigg(
                   - \beta_1 \overline {\cal G}_{P,1}^{~I}(\ep)
                   -4\beta_0 \overline {\cal G}_{P,2}^{~I}(\ep)\Bigg)
                  +\overline {\cal G}_{P,3}^{~I}(\ep)
\nonumber\\[2ex]
\overline {G}^{~I,(4)}_P(\ep)&=& {1 \over \ep^3} \Bigg(
                    -8 \beta_0^3 \overline {\cal G}_{P,1}^{~I}(\ep)\Bigg)
                  +{1 \over \ep^2} \Bigg(
                    {16 \over 3} \beta_0 \beta_1 \overline {\cal G}_{P,1}^{~I}(\ep)
                    +12\beta_0^2 \overline {\cal G}_{P,2}^{~I}(\ep)\Bigg)
\nonumber\\[2ex]
&&                  +{1 \over \ep} \Bigg(
                       -{2 \over 3} \beta_2 \overline {\cal G}_{P,1}^{~I}(\ep)
                       -2 \beta_1 \overline {\cal G}_{P,2}^{~I}(\ep)
                    -6 \beta_0 \overline {\cal G}_{P,3}^{~I}(\ep)\Bigg)
                  +\overline {\cal G}_{P,4}^{~I}(\ep)
\end{eqnarray}
The $z$ independent constants
$\overline {\cal G}^{~I}_{P,i}(\ep)$ 
are obtained by demanding the finiteness of $\Delta^{sv}_{~I,P}$ given in
eqn.(\ref{master}).   
Without setting $\ep=0$ in eqn.(\ref{master}), we expand $\Delta^{sv}_{~I,P}$ as
\begin{eqnarray}
\Delta^{sv}_{~I,P}(z,q^2,\mu_R^2,\mu_F^2,\ep)=\sum_{i=0}^\infty a_s^i(\mu_R^2)
\Delta^{sv,(i)}_{~I,P}(z,q^2,\mu_R^2,\mu_F^2,\ep)
\end{eqnarray}
Now, using the above expansion and eqn.(\ref{DYH}), we determine 
these constants by comparing the pole as well as non-pole terms
of the form factors, mass factorisation kernels and coefficient
functions $\Delta^{sv,(i-1)}_{~I,P}$ expanded in powers of $\ep$ to
the desired accuracy.  The factor $m$ appearing in the
solution turns out to be $1(1/2)$ for the DY and Higgs productions(DIS).  
This choice is governed by the number of
incoming partons that are responsible for the soft emissions.
Since $G^{~I}_P(\ep)$s in the form factors are found to satisfy 
a specific structure in terms of $B^I$, $f^I$,$\beta_i$ and $\gamma^b_i$
as given in eqn.(\ref{GI}),  
we find that the constants $\overline {\cal G}^{~I}_{P,i}(\ep)$
also satisfy similar looking expansion containing these constants.
\begin{eqnarray}
\overline {\cal G}^{~I}_{P,1}(\ep)&=&-\Big(f_1^I+B_1^I~\delta_{P,SJ}\Big)+
\sum_{k=1}^\infty \ep^k \overline {\cal G}^{~I,(k)}_{P,1}
\nonumber\\[2ex]
\overline {\cal G}^{~I}_{P,2}(\ep)&=&-\Big(f_2^I+B_2^I~\delta_{P,SJ}\Big)
-2 \beta_0 \overline{\cal G}_{P,1}^{~I,(1)}
+\sum_{k=1}^\infty\ep^k  \overline {\cal G}^{~I,(k)}_{P,2}
\nonumber\\[2ex]
\overline {\cal G}^{~I}_{P,3}(\ep)&=&-\Big(f_3^I+B_3^I~\delta_{P,SJ}\Big)
-2 \beta_1 \overline{\cal G}_{P,1}^{~I,(1)}
-2 \beta_0 \left(\overline{\cal G}_{P,2}^{~I,(1)}
+2 \beta_0 \overline{\cal G}_{P,1}^{~I,(2)}\right)
+\sum_{k=1}^\infty \ep^k \overline {\cal G}^{~I,(k)}_{P,3}
\nonumber \\[2ex]
\overline {\cal G}^{~I}_{P,4}(\ep)&=&
-\Big(f_4^I+B_4^I~\delta_{P,SJ}\Big)
-2 \beta_2 \overline {\cal G}^{~I,(1)}_{P,1}
-2 \beta_1 \Big( \overline {\cal G}^{~I,(1)}_{P,2} 
+ 4 \beta_0 \overline {\cal G}^{~I,(2)}_{P,1} \Big)
\nonumber\\[2ex]
&&-2 \beta_0 \Big(\overline {\cal G}^{~I,(1)}_{P,3}
+2 \beta_0 \overline {\cal G}^{~I,(2)}_{P,2}
                         +4 \beta_0^2 \overline {\cal G}^{~I,(3)}_{P,1}\Big)
        +\sum_{k=1}^\infty \ep^k \overline {\cal G}^{~I,(k)}_{P,4}
\label{OGI}
\end{eqnarray}
The above structure indicates that all the poles {\it including the
single pole} of the soft
distribution function can be predicted from that of the form factors,
renormalisation constants and the mass factorisation kernels.
This is possible because we have now better understanding
\cite{Ravindran:2004mb} of the
structure of even the single pole terms of the form factors.  
Notice that the terms proportional $\beta_i$ in the above
expansion are due to the coupling constant renormalisation.
The coefficients of single poles are proportional to
the cusp anomalous dimension $A^I$
and the combination $-(f^I+ B^I \delta_{P,SJ})$ 
where the constants $f_i^I$ and $B_i^I$ are process independent.
The $\ep$ dependent terms in $\overline {\cal G}^{~I}_P(\ep)$ can be obtained
from the fixed order(in $a_s$) computations of cross sections
and the finite parts of the form factors.  At the moment,
we know $\overline {\cal G}^{~I}_{P,1}(\ep)$ to all orders in $\ep$,
$\overline {\cal G}^{~I}_{P,2}(\ep)$ to order $\ep$ and $\overline 
{\cal G}^{~I}_{P,3}(\ep)$ to order ${\ep}^0$.
The lowest order term $\overline {\cal G}^{~I}_{P,1}(\ep)$
is known to all orders in $\ep$ because it is straight forward to
compute the fixed order soft contribution.  On the other hand,
it is technically hard to determine $\ep$ dependent parts 
of soft cross sections beyond the lowest order $a_s$.
We find,
\begin{eqnarray}
\overline{\cal G}^{~I,(1)}_{S,1}
&=&C_I~ \Big(-3 \zeta_2\Big)
\nonumber\\[2ex]
\overline{\cal G}^{~I,(2)}_{S,1}
&=& C_I~ \Bigg({7 \over 3}  \zeta_3\Bigg)
\nonumber\\[2ex]
\overline{\cal G}^{~I,(3)}_{S,1}
&=& C_I~ \Bigg(-{3 \over 16}  \zeta_2^2\Bigg)
\nonumber\\[2ex]
\overline{\cal G}^{~I,(1)}_{S,2}
&=& C_I C_A~ \Bigg({2428 \over 81} -{469 \over 9} \zeta_2
              +4 \zeta_2^2 -{176 \over 3} \zeta_3\Bigg)
\nonumber\\[2ex]
&&             +C_I n_f~ \Bigg(-{328 \over 81} + {70 \over 9} \zeta_2
                +{32 \over 3} \zeta_3 \Bigg)
\label{DIi}
\end{eqnarray}
where $C_I=C_F$ for $I=q,b$ and $C_I=C_A$ for $I=g$.
Interestingly these constants $\overline {\cal G}^{~I}_{S,i}(\ep)$ turn out to
be maximally non-abelien.  That is, they satisfy
\begin{eqnarray}
\overline {\cal G}^{~q}_{S,i}(\ep) = 
\overline {\cal G}^{~b}_{S,i}(\ep) = 
{C_F \over C_A}~ \overline {\cal G}^{~g}_{S,i}(\ep)
\end{eqnarray}
Similarly, for the DIS, the constants $\overline {\cal G}^q_{SJ,i}$ are 
found to be
\begin{eqnarray}
\overline{\cal G}^{~q,(1)}_{SJ,1}&=&
C_F~ \Big({7 \over 2}-3 \zeta_2\Big)
\nonumber\\[2ex]
\overline{\cal G}^{~q,(2)}_{SJ,1}&=&
C_F~ \Bigg(-{7 \over 2} +{9 \over 8} \zeta_2 +{7 \over 3}  \zeta_3\Bigg)
\nonumber\\[2ex]
\overline{\cal G}^{~q,(3)}_{SJ,1}&=&
C_F \Bigg( {7 \over 2} -{21 \over 16} \zeta_2
-{3 \over 16} \zeta_2^2 -{7 \over 8} \zeta_3\Bigg)
\nonumber\\[2ex]
\overline{\cal G}^{~q,(1)}_{SJ,2}&=&
C_F^2 \Bigg(
{9 \over 8} -{41 \over 2}\zeta_2
         +{82 \over 5} \zeta_2^2 -6 \zeta_3\Bigg)
\nonumber\\[2ex]
&&+C_F C_A~ \Bigg({69761 \over 648}-{1961 \over 36} \zeta_2
   -{17 \over 5} \zeta_2^2 -40 \zeta_3\Bigg)
\nonumber\\[2ex]
&&+ C_F n_f\Bigg(
-{5569 \over 324} +{163 \over 18} \zeta_2 +4 \zeta_3\Bigg)
\label{BIi}
\end{eqnarray}

The threshold corrections dominate when the partonic scaling 
variable $z$ approaches its kinematic limit which is $1$.  
They manifest in terms of the distributions
$\delta(1-z)$ and ${\cal D}_i$.  Since the hadronic
cross sections are expressed in terms of convolutions of partonic
cross sections and the parton distribution functions, it is
more convenient to study the threshold enhanced corrections in
Mellin $N$ space where $z\rightarrow 1$ corresponds to large $N$.
In Mellin $N$ space, all the convolutions
become ordinary products and the $\delta(1-z)$ distribution becomes 
a constant and the distributions ${\cal D}_j$ become functions of 
logarithm of $N$. 
The threshold resummation in Mellin $N$ space has been a successful
approach thanks to several important works given in 
\cite{Sterman:1986aj, Catani:1989ne, Contopanagos:1996nh, Eynck:2003fn}.
We show in the following how the soft distribution function 
$\Phi^I_P(\hat a_s,q^2,\mu^2,z,\ep)$
captures all the features of the $N$ space resummation approach. 
The exponents of the $z$ space resummed cross sections
get contributions from both form factor as well as the soft distribution
functions.  The form factor contributes to $\delta(1-z)$ part
and the soft distributions functions contribute to $\delta(1-z)$ as well as 
to the distributions ${\cal D}_j$.  Using
\begin{eqnarray}
{1 \over 1-z}\left[(1-z)^{2m}\right]^{i{\ep \over 2}}&=&
{1 \over m i \ep} \delta(1-z) +
\left({1 \over 1-z} \left[(1-z)^{2m}\right]^{i{\ep \over 2}} \right)_+
\end{eqnarray}
we can express the soft distribution function (for any $m$) as
\begin{eqnarray}
\Phi^{~I}_P(\hat a_s,q^2,\mu^2,z,\ep)&=&
\Bigg( {m \over 1-z} \Bigg\{
\int_{\mu_R^2}^{q^2 (1-z)^{2m}} {d \lambda^2 \over \lambda^2}
A_I \left(a_s(\lambda^2)\right) + \overline G^{~I}_P \left(
a_s\left(q^2 (1-z)^{2m}\right),\ep\right)\Bigg\} \Bigg)_+
\nonumber\\[2ex]
&&+\delta(1-z)~~~ \sum_{i=1}^\infty \hat a_s^i
\left({q^2 \over \mu^2}\right)^{i {\ep \over 2}}
S_{\ep}^i~
\hat \phi^{~I,(i)}_P(\ep)
\nonumber\\[2ex]
&&+\left({m \over 1-z}\right)_+ ~~~\sum_{i=1}^\infty
\hat a_s^i \left({\mu_R^2 \over \mu^2}\right)^{i {\ep \over 2}}
S_{\ep}^i~
\overline K^{~I,(i)}(\ep)
\label{resum}
\end{eqnarray}
where
\begin{eqnarray}
\overline G^{~I}_P\left(a_s \left(q^2 (1-z)^{2m} \right),\ep\right)
&=& \sum_{i=1}^\infty \hat a_s^i
\left({q^2 (1-z)^{2m} \over \mu^2}\right)^{i{\ep \over 2}}
S_{\ep}^i
\overline G^{~I,(i)}_P(\ep)
\label{Gbar2}
\end{eqnarray}
For $m=1$, that is, for DY and Higgs production, we can easily identify 
$G^{~I}_S\left(a_s \left(q^2 (1-z)^{2} \right),\ep\right)$
with the threshold exponent 
$D^I\left(a_s\left(q^2 (1-z)^{2}\right)\right)$:
\begin{eqnarray}
D^I\left(a_s\left(q^2 (1-z)^{2}\right)\right)
&=&\sum_{i=1}^\infty a_s^i\left(q^2 (1-z)^{2} \right) D_i^I
\nonumber\\[2ex]
&=&2~ \overline G^{~I}_S\left(a_s \left(q^2 (1-z)^{2} \right),\ep\right)
\Bigg|_{\ep=0}
\end{eqnarray}
Comparing the above result with the eqn.(\ref{Gbar1}), we find
\begin{eqnarray}
D^I_i = 2~\overline {\cal G}_{S,i}^{~I}(\ep=0)
\label{DI}
\end{eqnarray}
For $m=1/2$, that is, for DIS, we identify 
$G^{~I}_{SJ}\left(a_s \left(q^2 (1-z) \right),\ep\right)$
with the threshold exponent 
$B^I_{DIS}\left(a_s\left(q^2 (1-z)\right)\right)$:
\begin{eqnarray}
B^I_{DIS}\left(a_s\left(q^2 (1-z)\right)\right)
&=&\sum_{i=1}^\infty a_s^i\left(q^2 (1-z) \right) B_{DIS,i}^I
\nonumber\\[2ex]
&=&\overline G^{~I}_{SJ}\left(a_s \left(q^2 (1-z) \right),\ep\right)
\Bigg|_{\ep=0}
\end{eqnarray}
Comparing the above result with the eqn.(\ref{Gbar1}), we find
\begin{eqnarray}
B^I_{DIS,i} = \overline {\cal G}_{SJ,i}^{~I}(\ep=0)
\label{BI}
\end{eqnarray}
It is evident from the above derivation that the allowed value $m=1/2$ for
DIS does not allow an exponent similar to $D^I(a_s(q^2 (1-z)^2))$ that
appears in DY and Higgs productions.  
It is also straightforward to show that the so called 
conspicuous relation found in \cite{Moch:2005ba} follows 
directly from eqns.(\ref{OGI},\ref{DI},\ref{BI}):
\begin{eqnarray}
{1 \over 2} D^q_i - B^q_{DIS,i} =  B^q_i + \sum_{j=0}^{i-2}~ C^q_j(\beta) ~
\delta \overline {\cal G}^q_j
\end{eqnarray} 
where the constants $C^q_j(\beta)$ and 
$\delta \overline {\cal G}^q_i$ can be obtained from the 
eqns.(\ref{OGI},\ref{DIi},\ref{BIi}).

Few important remarks about the eqn.(\ref{resum}) are in order.
The third line in the eqn.(\ref{resum}) cancels against 
${\cal D}_0$ part of the mass factorisation kernel.
The second line contains the right poles
in $\ep$ to cancel those coming from the form factor as well as
the mass factorisation kernel.  In addition, it 
contains the terms that are finite as $\ep$ becomes zero through
the constants $\overline {\cal G}^{~I}_{P,i}(\ep)$.  These finite constants
contribute to soft part of the cross section at higher orders.    
Hence, adding the eqn.(\ref{resum}) with the renormalised form factor and the
mass factorisation kernels, performing the coupling constant renormalisation
and then finally taking Mellin moment, 
we reproduce the resummed result given in
\cite{Sterman:1986aj, Catani:1989ne, Contopanagos:1996nh, Eynck:2003fn}
when $\ep \rightarrow 0$.  

In \cite{Ravindran:2005vv}, it was observed that the soft distribution functions 
$\Phi^{~I}_P(\hat a_s,q^2,\mu^2,z,\ep)$
of DY and Higgs production are
maximally non-abelien and hence the entire soft contribution of Higgs production
can be predicted from that of DY and vice versa.  
In fact the $\ep$ 
dependent parts of $\overline {\cal G}^q_{S,i}(\ep)$ 
extracted from DY are needed to predict the soft part of the Higgs 
production.   In addition, using the resummed result 
given in eqn.(\ref{master}), and the available
exponents $g_i^{~I}(\ep),\overline {\cal G}_i^{~I}(\ep)$, 
one can also predict part of the higher order soft plus 
virtual contributions to the cross sections.  The available
exponents are 
\begin{eqnarray}
&g_1^{~I,j}~,~~~
\overline {\cal G}_1^{~I,(j)} \hspace{1.5cm} {\rm for}\hspace{1.5cm}  j={\rm all}
\nonumber\\[2ex]
&g_2^{~I,j}~,~~~\overline {\cal G}_2^{~I,(j)} 
\hspace{1.5cm} {\rm for} \hspace{1.5cm} j=0,1
\nonumber\\[2ex]
&g_3^{~I,j}~,~~~\overline {\cal G}_3^{~I,(j)} 
\hspace{1.5cm} {\rm for} \hspace{1.5cm} j=0
\nonumber
\end{eqnarray}
in addition to the known $\beta_i~(i=0,1,2,3)$, the constants in the splitting
functions $A_i,~ B_i~~(i=1,2,3)$, the maximally non-abelien constants
$f_i~(i=1,2,3)$ and the anomalous dimensions $\gamma^b_i(i=0,1,2,3)$.  For
$I=q,g$, the constants $g^{q,j}_2$ and $g^{g,j}_2$ are known for $j=2,3$ also
(see \cite{Moch:2005id}).  

Using the resummed expression given in eqn.(\ref{master}) and the known
exponents, we present here the results for $\Delta_{I,P}^{sv,(i)}$ for
DY, Higgs production and DIS.  As we have already mentioned in the beginning, for DIS, 
it is just a consistency check
because the three loop form factors were derived from the
DIS coefficient functions.  The Drell-Yan coefficients 
$\Delta_{q,S}^{sv,(i)}$ and 
Higgs coefficients $\Delta_{g,S}^{sv,(i)}$ and $\Delta_{b,S}^{sv,(i)}$
are known upto NNLO ($i=0,1,2$) 
from the explicit computations
(see \cite{Altarelli:1978id}-\cite{Harlander:2003ai}).
For $N^3LO$ for $I=q,g$, {\it a partial result} $\Delta_{I,S}^{sv,(3)}$, 
i.e., a result without 
$\delta(1-z)$ part is available from the work of 
\cite{Moch:2005ky}.  Using the universal behavior
of the soft distribution function for $P=S$, in the reference
\cite{Ravindran:2005vv}, the 
entire $\Delta_{I,S}^{sv,(i)}$ for $I=q,g$ upto NNLO 
as well as {\it partial} $N^3LO$ for Drell-Yan and Higgs through gluon fusion, 
that is the coefficients $\Delta_{I,S}^{sv,(3)}$ for $I=q,g$ were reproduced.
This was achieved using the resummation formula given in eqn.(\ref{master}).  
In this article,  we predict for the first time the $N^3LO$ 
contribution $\Delta_{b,S}^{sv,(3)}$ to Higgs production through 
bottom quark annihilation with the same accuracy that 
the coefficients $\Delta_{I,S}^{sv,(3)}$ for $I=q,g$
are known.  In addition, we extend this approach  
to $N^4LO$ order where we can predict 
{\it partial} soft plus virtual
contribution coming from all ${\cal D}_j$ except $j=0,1$ for
Drell-Yan $N^4LO$ coefficient $\Delta_{q,S}^{sv,(4)}$, 
gluon fusion to Higgs $N^4LO$ coefficient $\Delta_{g,S}^{sv,(4)}$ and
bottom quark annihilation to Higgs boson $N^4LO$ coefficient
$\Delta_{b,S}^{sv,(4)}$.  Like $N^3LO$
$\Delta_{I,S}^{sv,(3)}$, here also we can not predict $\delta(1-z)$ part.   
These results are presented in the
Appendix B for $\mu_R^2=\mu_F^2=q^2$.  
Using our resummed formula
and the DIS exponents ${\cal G}_{SJ}^{q}(\ep)$, we have
reproduced\footnote{after fixing the typos in eqn.(4.19) of \cite{Vermaseren:2005qc}
and in eqns.(5.6,5.7) and eqn.(5.9) of \cite{Moch:2005ba}} the partial 
$N^4LO$ and $N^3LO$ coefficients $\Delta_{q,SJ}^{sv,(4)}$, 
$\Delta_{q,SJ}^{sv,(3)}$ for DIS
given in \cite{Moch:2005ba,Vermaseren:2005qc} and the lower order results 
(see \cite{vanNeerven:1991nn} for $NNLO$). 
The computation of these soft plus virtual contributions to DY, 
Higgs productions and DIS involves
convolutions of distribution functions ${\cal D}_j$ with $j=0,...,7$.  
We have computed them using the following integral representation given in 
\cite{Anastasiou:2005qj}.
\begin{eqnarray}
{\cal D}_i \otimes {\cal D}_j &=&\lim_{\ep \rightarrow 0}~~ \lim_{a,b\rightarrow 1}
\Bigg({\partial \over \partial a}\Bigg)^i
\Bigg({\partial \over \partial b}\Bigg)^j {1 \over \ep^{i+j}}~
~~\int_0^1 dx \int_0^1 dy~ (1-x)^{a ~\ep -1} (1-y)^{b~ \ep -1}
\nonumber\\[2ex]
&&\times \Bigg\{ \delta(z-x y) -\delta(z-x) -\delta(z-y)+\delta(1-z)\Bigg\}
\end{eqnarray}
The above integral reduces to a set of Hypergeometric functions and Euler
Gamma functions which can be expanded around $\ep=0$ to desired accuracy.  
We are interested only in those terms which are proportional to 
the distributions $\delta(1-z)$ and ${\cal D}_i$ for our analysis.  
For this purpose, we have derived a formula to compute
the convolutions of distributions for any arbitrary $i,j$ using the above
integral representation:
\begin{eqnarray}
{\cal D}_i \otimes {\cal D}_j &=&\lim_{\ep \rightarrow 0}~~ \lim_{a,b\rightarrow 1}
\Bigg({\partial \over \partial a}\Bigg)^i
\Bigg({\partial \over \partial b}\Bigg)^j {1 \over \ep^{i+j+1}}~
\Bigg({a+b \over ab}\Bigg)
\Bigg(
~~\Bigg[{\Gamma(1+a\ep) \Gamma(1+b\ep) \over \Gamma(1+(a+b)\ep)}-1\Bigg]
\nonumber\\[2ex]&&
\times \Bigg[{1 \over (a+b)~ \ep} \delta(1-z)
+\sum_{k=0}^\infty { \Big((a+b)\ep\Big)^k \over k!} {\cal D}_k\Bigg]
+\sum_{k=1}^\infty { \Big((a+b)\ep \Big)^k \over k!} {\cal D}_k 
\nonumber\\[2ex]&&
-~{a \over a+b}~\sum_{k=1}^\infty { \big(a\ep \big)^k \over k!} {\cal D}_k 
-~{b \over a+b}~\sum_{k=1}^\infty { \big(b\ep \big)^k \over k!} {\cal D}_k 
+R(a,b,z,\ep)\Bigg)
\end{eqnarray}
where $R(a,b,z,\ep)$ is the remaining regular function.
It is now straightforward to obtain 
$\Delta_{I,S}^{sv,(i)}$ for $i=1,...,4$ for both Higgs($I=g,b$) and DY($I=q$)
productions.  

The impact of partial soft plus virtual parts of $N^3LO$ and $N^4LO$ 
contributions to Higgs production through gluon fusion 
at LHC is presented in figure (1).
The Higgs production cross section is given by
\begin{eqnarray}
\sigma^H(S,m_H^2)={\pi G_B^2 \over 8 (N^2-1)}
\sum_{a,b=q,\overline q,g} \int_x^1~ dy~ \Phi_{ab}(y,\mu_F^2) 
~\Delta^H_{ab}\left({x \over y},m_H^2,\mu_F^2,\mu_R^2\right)
\end{eqnarray}  
where $x=m_H^2/S$, $N=3$ and the factor $G_B$ can be found from
\cite{Chetyrkin:1997un}.  The flux $\Phi_{ab}(y,\mu_F^2)$ is
given by
\begin{eqnarray}
\Phi_{ab}(y,\mu_F^2)=\int_y^1 {dw \over w}~ f_a(w,\mu_F^2) 
~f_b\left({y \over w},\mu_F^2\right)
\end{eqnarray}
where $f_a(w,\mu_F^2)$ is the parton distribution function.
The partonic cross section $\Delta^H_{ab}$ contains both
soft plus virtual as well as hard contributions:
\begin{eqnarray}
\Delta^H_{ab}(z,m_H^2,\mu_R^2,\mu_F^2)=\Delta^{sv}_{g,S}(z,m_H^2,\mu_R^2,\mu_F^2)
+\Delta^{H,hard}_{ab}(z,m_H^2,\mu_R^2,\mu_F^2)
\end{eqnarray}

\begin{figure}[htb]
\vspace{1mm}
\centerline{
\epsfig{file=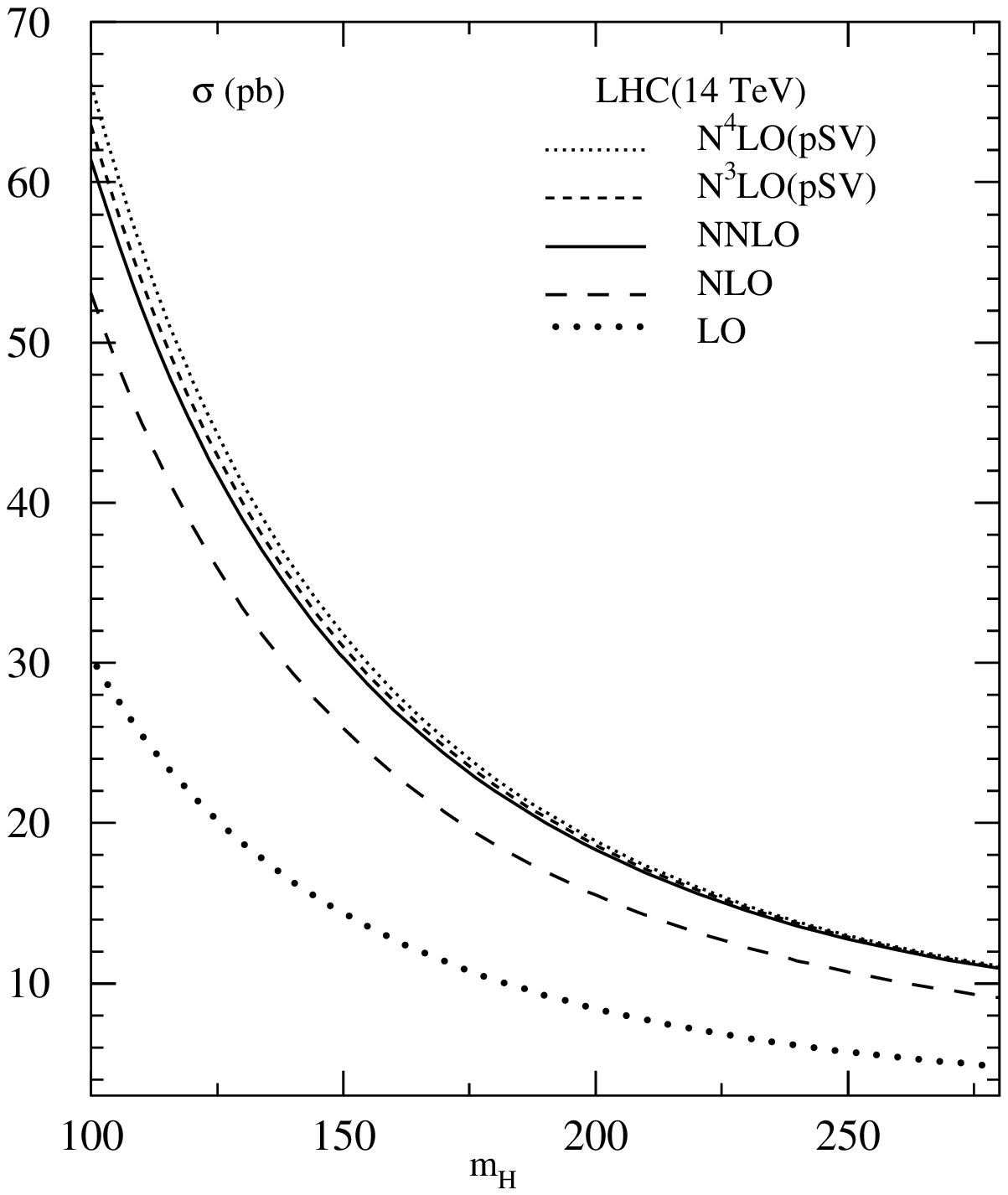,width=8cm,height=10cm,angle=0}
\epsfig{file=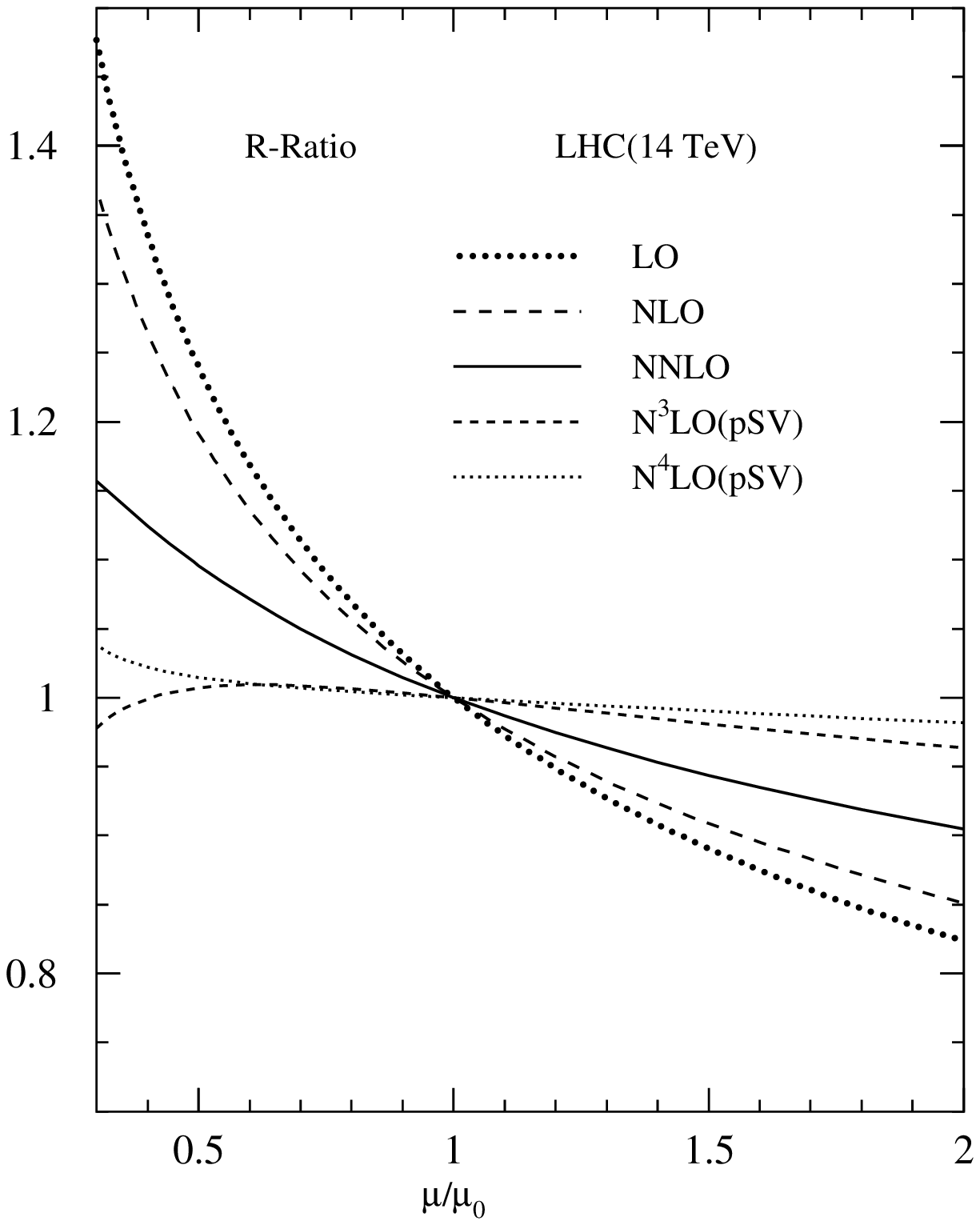,width=8cm,height=9.3cm,angle=0}
}
\caption{
Total cross section for Higgs production through gluon fusion at LHC,
and its scale dependence for the Higgs boson of mass $m_H=150~GeV$ with
$\mu=m_R,\mu_0=\mu_H$. The abbreviation "pSV" means partial soft plus virtual.
}
\end{figure}

We choose the center of mass energy to be $\sqrt{S}=14~TeV$ for LHC.
The standard model parameters that enter our computation
are the Fermi constant 
$G_F=4541.68$, $Z$ boson mass $M_Z=91.1876~GeV$,
top quark mass $m_t=173.4~GeV$.  The strong coupling constant
$\alpha_s(\mu_R^2)$ is evolved using 4-loop RG equations
depending on the order in which the cross section is evaluated.
We choose $\alpha^{LO}_s(M_Z)=0.130$, $\alpha^{NLO}_s(M_Z)=0.119$,
$\alpha^{NNLO}_s(M_Z)=0.115$ and $\alpha^{N^iLO}_s(M_Z)=0.114$ for $i>2$.
We use MRST 2001 LO for leading order, MRST2001 NLO for next to leading order
and MRST 2002 NNLO for $N^iLO$ with $i>1$ \cite{Martin:2002dr,Martin:2001es}.
In the first plot of fig.(1), we have shown the cross sections in $pb$ for various
values of Higgs mass.  For LO,NLO and NNLO we used the exact
results which contain both soft plus virtual as well as regular hard
contributions.  For $N^iLO$ (i=3,4), we use only soft plus virtual
results extracted from the resummed formula.  Here we have set
$\mu_F=\mu_R=m_H$.  We find that the inclusion of $N^iLO$($i=3,4$)
does not change the cross section much confirming the reliability
of the perturbative approach.   In the second plot of fig.(1), we have plotted 
the scale variation of the cross section using the ratio:
\begin{eqnarray}
R\left({\mu_R^2 \over m_H^2}\right)=
{\sigma^H\left(S,m_H^2,\mu_R^2,\mu_F^2=m_H^2\right) 
\over \sigma^H\left(S,m_H^2,\mu_R^2=\mu_F^2=m_H^2\right)} 
\end{eqnarray} 
It is clear from the second plot of fig.(1) that the inclusion of 
$N^iLO$($i=3,4$) soft plus virtual contributions
reduces the scale ambiguity further.  

To summarise, we have systematically studied the soft plus virtual
correction to inclusive processes such as
DIS, DY, Higgs productions through gluon fusion and bottom quark annihilation.  
Using renormalisation group invariance and  
Sudakov resummation of scattering amplitudes and the factorisation
property of these hard scattering cross sections, the resummation of 
these corrections has been achieved.  We have also shown 
how these resummed distributions
are related to resummation exponents that appear in Mellin space.
Using our resummed results we predict partial soft plus virtual cross sections
at $N^3LO$ and $N^4LO$.  We have also shown the phenomenological consequences
of such results for Higgs production through gluon fusion at LHC. 

Acknowledgments:  The author would like to thank Prakash Mathews for discussion. 

\myappendix
\section{Appendix}
In this appendix, we present the logarithm of the form factor 
$\ln \hat {\cal F}^b$ for the Yukawa interaction
of Higgs boson with bottom quarks upto three loop level expanded
in powers of $\ep$ to the desired accuracy for our computation. 
\begin{eqnarray}
\ln \hat {\cal F}^b(\hat a_s,Q^2,\mu^2,\ep)&\!\!\!\!\! =\!\!\!\!\!&
\hat a_s \left({Q^2 \over \mu^2}\right)^{{\ep \over 2}} S_{\ep} C_F \Bigg[
        {1\over \ep^2}    \Big(
          - 8
          \Big)
       + \ep    \Bigg(
           2
          - {7\over 3}  \zeta_3
          \Bigg)
       + \ep^2    \Bigg(
          - 2
          + {1\over 4}  \zeta_2
          + {47\over 80}  \zeta_2^2
          \Bigg)
\nonumber\\[2ex]
&&       +  \Bigg(
          - 2
          + \zeta_2
          \Bigg)
\Bigg]
\nonumber\\[2ex]
&& +\hat a_s^2 \left({Q^2 \over \mu^2}\right)^{{\ep}} S^2_{\ep}\Bigg[
{1\over \ep^3} \Bigg( n_f  C_F    \Bigg(
          - {8\over 3}
          \Bigg)
       +   C_F  C_A    \Bigg(
           {44\over 3}
          \Bigg)
\Bigg)
\nonumber\\[2ex]
&&+ {1\over \ep^2} \Bigg( n_f  C_F    \Bigg(
           {20\over 9}
          \Bigg)
       +   C_F  C_A    \Bigg(
          - {134\over 9}
          + 4  \zeta_2
          \Bigg)
\Bigg)
+ {1\over \ep}  \Bigg( n_f  C_F    \Bigg(
          - {92\over 27}
          - {2\over 3}  \zeta_2
          \Bigg)
\nonumber\\[2ex]
&&       +   C_F  C_A    \Bigg(
           {440\over 27}
          + {11\over 3}  \zeta_2
          - 26  \zeta_3
          \Bigg)
       +  C_F^2    \Bigg(
          - 12  \zeta_2
          + 24  \zeta_3
          \Bigg)
\Bigg)
       + n_f  C_F    \Bigg(
           {416\over 81}
\nonumber\\[2ex]
&&          + {5\over 9}  \zeta_2
          - {26\over 9}  \zeta_3
          \Bigg)
       + C_F  C_A    \Bigg(
          - {1655\over 81}
          - {103\over 18}  \zeta_2
          + {44\over 5}  \zeta_2^2
          + {305\over 9}  \zeta_3
          \Bigg)
\nonumber\\[2ex]
&&       + C_F^2    \Bigg(
          4
          + 16  \zeta_2
          - {44\over 5}  \zeta_2^2
          - 30  \zeta_3
          \Bigg)
\Bigg]
\nonumber\\[2ex]
&&+\hat a_s^3\left({Q^2 \over \mu^2}\right)^{3{\ep \over 2}}S^3_{\ep} \Bigg[
{1\over \ep^4} \Bigg(  n_f  C_F  C_A    \Bigg(
           {1408\over 81}
          \Bigg)
       +   n_f^2  C_F    \Bigg(
          - {128\over 81}
          \Bigg)
\nonumber\\[2ex]
&&       +   C_F  C_A^2    \Bigg(
          - {3872\over 81}
          \Bigg)
\Bigg)
+ {1\over \ep^3} \Bigg( n_f  C_F  C_A    \Bigg(
          - {8528\over 243}
          + {128\over 27}  \zeta_2
          \Bigg)
       +  n_f  C_F^2    \Bigg(
          - {16\over 9}
          \Bigg)
\nonumber\\[2ex]
&&       +  n_f^2  C_F    \Bigg(
           {640\over 243}
          \Bigg)
       +  C_F  C_A^2    \Bigg(
           {26032\over 243}
          - {704\over 27}  \zeta_2
          \Bigg)
\Bigg)
+ {1\over \ep^2}  \Bigg( n_f  C_F  C_A    \Bigg(
           {13640\over 243}
\nonumber\\[2ex]
&&          + {1264\over 81}  \zeta_2
          - {1024\over 27}  \zeta_3
          \Bigg)
       +  n_f  C_F^2    \Bigg(
           {220\over 27}
          - {64\over 3}  \zeta_2
          + {320\over 9}  \zeta_3
          \Bigg)
\nonumber\\[2ex]
&&       +  n_f^2  C_F    \Bigg(
          - {128\over 27}
          - {16\over 9}  \zeta_2
          \Bigg)
       +   C_F  C_A^2    \Bigg(
          - {38828\over 243}
          - {2212\over 81}  \zeta_2
          - {352\over 45}  \zeta_2^2
\nonumber\\[2ex]
&&          + {6688\over 27}  \zeta_3
          \Bigg)
       +   C_F^2  C_A    \Bigg(
           {352\over 3}  \zeta_2
          - {704\over 3}  \zeta_3
          \Bigg)
\Bigg)
+ {1\over \ep}  \Bigg(n_f  C_F  C_A    \Bigg(
          - {180610\over 2187}
\nonumber\\[2ex]
&&          - {10066\over 243}  \zeta_2
          + {88\over 5}  \zeta_2^2
          + {10280\over 81}  \zeta_3
          \Bigg)
       +  n_f  C_F^2    \Bigg(
          - {1171\over 81}
          + {454\over 9}  \zeta_2
          - {496\over 45}  \zeta_2^2
\nonumber\\[2ex]
&&          - {3104\over 27}  \zeta_3
          \Bigg)
       +  n_f^2  C_F    \Bigg(
           {19232\over 2187}
          + {80\over 27}  \zeta_2
          - {272\over 81}  \zeta_3
          \Bigg)
       +  C_F  C_A^2    \Bigg(
           {385325\over 2187}
\nonumber\\[2ex]
&&          + {176\over 9}  \zeta_2  \zeta_3
          + {31966\over 243}  \zeta_2
          - {1604\over 15}  \zeta_2^2
          - {17084\over 27}  \zeta_3
          + {272\over 3}  \zeta_5
          \Bigg)
       +  C_F^2  C_A    \Bigg(
          - 12
\nonumber\\[2ex]
&&          + {32\over 3}  \zeta_2  \zeta_3
          - {2932\over 9}  \zeta_2
          + {3832\over 45}  \zeta_2^2
          + {5648\over 9}  \zeta_3
          + 80  \zeta_5
          \Bigg)
       +  C_F^3    \Bigg(
          - {100\over 3}
\nonumber\\[2ex]
&&          - {64\over 3}  \zeta_2  \zeta_3
          + 12  \zeta_2
          + {192\over 5}  \zeta_2^2
          + {136\over 3}  \zeta_3
          - 160  \zeta_5
          \Bigg)
\Bigg)
\Bigg]
\end{eqnarray}
\section{Appendix}
In this appendix we present soft plus virtual parts
of Drell-Yan production of di-leptons 
and Higgs productions through bottom quark annihilation and gluon fusion.
\begin{eqnarray}
\Delta_{q,S}^{sv,(0)} &=& \delta(1-z)
\\[2ex]
\Delta_{q,S}^{sv,(1)} &=& 
C_F \Bigg(
        \delta(1-z)    \Bigg[
          - 16
          + 8  \zeta_2
          \Bigg]
       + {\cal D}_1    \Bigg[
           16
          \Bigg]
\Bigg)
\\[2ex]
\Delta_{q,S}^{sv,(2)}&=&
\delta(1-z)  \Bigg[
          n_f  C_F    \Bigg(
           {127\over  6}
          - {112\over  9}  \zeta_2
          + 8  \zeta_3
          \Bigg)
       +   C_F  C_A    \Bigg(
          - {1535\over  12}
          + {592\over  9}  \zeta_2
          - {12\over  5}  \zeta_2^2
          + 28  \zeta_3
          \Bigg)
\nonumber\\[2ex]
&&       +   C_F^2    \Bigg(
           {511\over  4}
          - 70  \zeta_2
          + {8\over  5}  \zeta_2^2
          - 60  \zeta_3
          \Bigg)
\Bigg]
+ {\cal D}_0 \Bigg[ 
          n_f  C_F    \Bigg(
           {224\over  27}
          - {32\over  3}  \zeta_2
          \Bigg)
       +   C_F  C_A    \Bigg(
          - {1616\over  27}
\nonumber\\[2ex]
&&          + {176\over  3}  \zeta_2
          + 56  \zeta_3
          \Bigg)
       +   C_F^2    \Bigg(
           256  \zeta_3
          \Bigg)
\Bigg]
+ {\cal D}_1  \Bigg[
         n_f  C_F    \Bigg(
          - {160\over  9}
          \Bigg)
       +   C_F  C_A    \Bigg(
           {1072\over  9}
          - 32  \zeta_2
          \Bigg)
\nonumber\\[2ex]
&&       +   C_F^2    \Bigg(
          - 256
          - 128  \zeta_2
          \Bigg)
\Bigg]
+ {\cal D}_2  \Bigg[  
        n_f  C_F    \Bigg(
           {32\over  3}
          \Bigg)
       +   C_F  C_A    \Bigg(
          - {176\over  3}
          \Bigg)
\Bigg]
+ {\cal D}_3  C_F^2    \Bigg[
           128
          \Bigg]
\\[2ex]
\Delta_{q,S}^{sv,(3)}&=&
{\cal D}_0 \Bigg[ 
          n_f  C_F  C_A    \Bigg(
           {125252\over  729}
          - {29392\over  81}  \zeta_2
          + {736\over  15}  \zeta_2^2
          - {2480\over  9}  \zeta_3
          \Bigg)
       +   n_f  C_F^2    \Bigg(
          - 6
\nonumber\\[2ex]
&&          + {1952\over  27}  \zeta_2
          - {1472\over  15}  \zeta_2^2
          - {5728\over  9}  \zeta_3
          \Bigg)
       +   n_f^2  C_F    \Bigg(
          - {3712\over  729}
          + {640\over  27}  \zeta_2
          + {320\over  27}  \zeta_3
          \Bigg)
\nonumber\\[2ex]
&&       +   C_F  C_A^2    \Bigg(
          - {594058\over  729}
          - {352\over  3}  \zeta_2  \zeta_3
          + {98224\over  81}  \zeta_2
          - {2992\over  15}  \zeta_2^2
          + {40144\over  27}  \zeta_3
          - 384  \zeta_5
          \Bigg)
\nonumber\\[2ex]
&&       +   C_F^2  C_A    \Bigg(
           {25856\over  27}
          - 1472  \zeta_2  \zeta_3
          - {12416\over  27}  \zeta_2
          + {1408\over  3}  \zeta_2^2
          + {26240\over  9}  \zeta_3
          \Bigg)
\nonumber\\[2ex]
&&       +   C_F^3    \Bigg(
          - 6144  \zeta_2  \zeta_3
          - 4096  \zeta_3
          + 12288  \zeta_5
          \Bigg)
\Bigg]
\nonumber\\[2ex]
&&+ {\cal D}_1  \Bigg[
         n_f  C_F  C_A    \Bigg(
          - {32816\over  81}
          + 384  \zeta_2
          \Bigg)
       +   n_f  C_F^2    \Bigg(
           {4288\over  9}
          + {2048\over  9}  \zeta_2
          + 1280  \zeta_3
          \Bigg)
\nonumber\\[2ex]
&&       +   n_f^2  C_F    \Bigg(
           {1600\over  81}
          - {256\over  9}  \zeta_2
          \Bigg)
       +   C_F  C_A^2    \Bigg(
           {124024\over  81}
          - {12032\over  9}  \zeta_2
          + {704\over  5}  \zeta_2^2
          - 704  \zeta_3
          \Bigg)
\nonumber\\[2ex]
&&       +   C_F^2  C_A    \Bigg(
          - {35572\over  9}
          - {11648\over  9}  \zeta_2
          + {3648\over  5}  \zeta_2^2
          - 5184  \zeta_3
          \Bigg)
       +   C_F^3    \Bigg(
           2044
          + 2976  \zeta_2
\nonumber\\[2ex]
&&          - {14208\over  5}  \zeta_2^2
          - 960  \zeta_3
          \Bigg)
\Bigg]
+ {\cal D}_2  \Bigg[
           n_f  C_F  C_A    \Bigg(
           {9248\over  27}
          - {128\over  3}  \zeta_2
          \Bigg)
       +   n_f  C_F^2    \Bigg(
           {544\over  9}
\nonumber\\[2ex]
&&          - {2048\over  3}  \zeta_2
          \Bigg)
       +   n_f^2  C_F    \Bigg(
          - {640\over  27}
          \Bigg)
       +   C_F  C_A^2    \Bigg(
          - {28480\over  27}
          + {704\over  3}  \zeta_2
          \Bigg)
       +   C_F^2  C_A    \Bigg(
          - {4480\over  9}
\nonumber\\[2ex]
&&          + {11264\over  3}  \zeta_2
          + 1344  \zeta_3
          \Bigg)
       +   C_F^3    \Bigg(
           10240  \zeta_3
          \Bigg)
\Bigg]
+ {\cal D}_3  \Bigg[ 
          n_f  C_F  C_A    \Bigg(
          - {2816\over  27}
          \Bigg)
\nonumber\\[2ex]
&&       +   n_f  C_F^2    \Bigg(
          - {2560\over  9}
          \Bigg)
       +   n_f^2  C_F    \Bigg(
           {256\over  27}
          \Bigg)
       +   C_F  C_A^2    \Bigg(
           {7744\over  27}
          \Bigg)
       +  C_F^2  C_A    \Bigg(
           {17152\over  9}
          - 512  \zeta_2
          \Bigg)
\nonumber\\[2ex]
&&       +   C_F^3    \Bigg(
          - 2048
          - 3072  \zeta_2
          \Bigg)
\Bigg]
+ {\cal D}_4  \Bigg[
           n_f  C_F^2    \Bigg(
           {1280\over  9}
          \Bigg)
       + {\cal D}_4  C_F^2  C_A    \Bigg(
          - {7040\over  9}
          \Bigg)
\Bigg]
\nonumber\\[2ex]
&&+ {\cal D}_5  C_F^3    \Bigg[
           512
          \Bigg]
\\[2ex]
\Delta_{q,S}^{sv,(4)}&=&
{\cal D}_2 \Bigg[ 
           n_f  C_F  C_A^2    \Bigg(
           {82216\over  9}
          - {64000\over  9}  \zeta_2
          + {1408\over  5}  \zeta_2^2
          - 1408  \zeta_3
          \Bigg)
       +   n_f  C_F^2  C_A    \Bigg(
           {337120\over  243}
\nonumber\\[2ex]
&&          - 31232  \zeta_2
          + {13184\over  3}  \zeta_2^2
          - {334976\over  9}  \zeta_3
          \Bigg)
       +   n_f  C_F^3    \Bigg(
           {2072\over  3}
          + {14144\over  3}  \zeta_2
          - {84224\over  15}  \zeta_2^2
\nonumber\\[2ex]
&&          - {109184\over  3}  \zeta_3
          \Bigg)
       +   n_f^2  C_F  C_A    \Bigg(
          - {31576\over  27}
          + {9728\over  9}  \zeta_2
          \Bigg)
       +   n_f^2  C_F^2    \Bigg(
          - {29744\over  243}
          + {60416\over  27}  \zeta_2
\nonumber\\[2ex]
&&          + {28160\over  9}  \zeta_3
          \Bigg)
       +   n_f^3  C_F    \Bigg(
           {3200\over  81}
          - {512\over  9}  \zeta_2
          \Bigg)
       +   C_F  C_A^3    \Bigg(
          - {1651520\over  81}
          + {138880\over  9}  \zeta_2
\nonumber\\[2ex]
&&          - {7744\over  5}  \zeta_2^2
          + 7744  \zeta_3
          \Bigg)
       +   C_F^2  C_A^2    \Bigg(
          - {1426292\over  243}
          - 5504  \zeta_2  \zeta_3
          + {2646016\over  27}  \zeta_2
          - {337216\over  15}  \zeta_2^2
\nonumber\\[2ex]
&&          + {359296\over  3}  \zeta_3
          - 9216  \zeta_5
          \Bigg)
       +   C_F^3  C_A    \Bigg(
           {139396\over  9}
          - 93696  \zeta_2  \zeta_3
          - {87008\over  3}  \zeta_2
          + {437888\over  15}  \zeta_2^2
\nonumber\\[2ex]
&&          + {632128\over  3}  \zeta_3
          \Bigg)
       +   C_F^4    \Bigg(
          - 409600  \zeta_2  \zeta_3
          - 163840  \zeta_3
          + 688128  \zeta_5
          \Bigg)
\Bigg]
\nonumber\\[2ex]
&&+ {\cal D}_3  \Bigg[
           n_f  C_F  C_A^2    \Bigg(
          - {117184\over  27}
          + {5632\over  9}  \zeta_2
          \Bigg)
       +   n_f  C_F^2  C_A    \Bigg(
          - {2143808\over  243}
          + {50176\over  3}  \zeta_2
\nonumber\\[2ex]
&&          + {7168\over  9}  \zeta_3
          \Bigg)
       +   n_f  C_F^3    \Bigg(
           {44224\over  9}
          + {88064\over  9}  \zeta_2
          + {306176\over  9}  \zeta_3
          \Bigg)
       +   n_f^2  C_F  C_A    \Bigg(
           {18304\over  27}
          - {512\over  9}  \zeta_2
          \Bigg)
\nonumber\\[2ex]
&&       +   n_f^2  C_F^2    \Bigg(
           {124288\over  243}
          - {4096\over  3}  \zeta_2
          \Bigg)
       +   n_f^3  C_F    \Bigg(
          - {2560\over  81}
          \Bigg)
       +   C_F  C_A^3    \Bigg(
           {698368\over  81}
          - {15488\over  9}  \zeta_2
          \Bigg)
\nonumber\\[2ex]
&&       +   C_F^2  C_A^2    \Bigg(
           {7699456\over  243}
          - 52736  \zeta_2
          + {13824\over  5}  \zeta_2^2
          - {140800\over  9}  \zeta_3
          \Bigg)
       +   C_F^3  C_A    \Bigg(
          - {421792\over  9}
\nonumber\\[2ex]
&&          - {536576\over  9}  \zeta_2
          + {100864\over  5}  \zeta_2^2
          - {1499648\over  9}  \zeta_3
          \Bigg)
       +   C_F^4    \Bigg(
           16352
          + 56576  \zeta_2
          - {195584\over  5}  \zeta_2^2
\nonumber\\[2ex]
&&          - 7680  \zeta_3
          \Bigg)
\Bigg]
+ {\cal D}_4  \Bigg[  
           n_f  C_F  C_A^2    \Bigg(
           {7744\over  9}
          \Bigg)
       +   n_f  C_F^2  C_A    \Bigg(
           {175360\over  27}
          - {2560\over  3}  \zeta_2
          \Bigg)
\nonumber\\[2ex]
&&       +   n_f  C_F^3    \Bigg(
          - {14080\over  27}
          - {87040\over  9}  \zeta_2
          \Bigg)
       +   n_f^2  C_F  C_A    \Bigg(
          - {1408\over  9}
          \Bigg)
       +   n_f^2  C_F^2    \Bigg(
          - {12800\over  27}
          \Bigg)
\nonumber\\[2ex]
&&       +   n_f^3  C_F    \Bigg(
           {256\over  27}
          \Bigg)
       +   C_F  C_A^3    \Bigg(
          - {42592\over  27}
          \Bigg)
       +   C_F^2  C_A^2    \Bigg(
          - {536960\over  27}
          + {14080\over  3}  \zeta_2
          \Bigg)
\nonumber\\[2ex]
&&       +   C_F^3  C_A    \Bigg(
           {79360\over  27}
          + {478720\over  9}  \zeta_2
          + 8960  \zeta_3
          \Bigg)
       +   C_F^4    \Bigg(
           {286720\over  3}  \zeta_3
          \Bigg)
\Bigg]
\nonumber\\[2ex]
&&+ {\cal D}_5  \Bigg[
           n_f  C_F^2  C_A    \Bigg(
          - {45056\over  27}
          \Bigg)
       +   n_f  C_F^3    \Bigg(
          - {5120\over  3}
          \Bigg)
       +   n_f^2  C_F^2    \Bigg(
           {4096\over  27}
          \Bigg)
\nonumber\\[2ex]
&&       +   C_F^2  C_A^2    \Bigg(
           {123904\over  27}
          \Bigg)
       +   C_F^3  C_A    \Bigg(
           {34304\over  3}
          - 3072  \zeta_2
          \Bigg)
       +   C_F^4    \Bigg(
          - 8192
          - 20480  \zeta_2
          \Bigg)
\Bigg]
\nonumber\\[2ex]
&&+ {\cal D}_6  \Bigg[
          n_f  C_F^3    \Bigg(
           {7168\over  9}
          \Bigg)
       +  C_F^3  C_A    \Bigg(
          - {39424\over  9}
          \Bigg)
\Bigg]
+ {\cal D}_7  
            C_F^4    \Bigg[
           {4096\over  3}
          \Bigg]
\end{eqnarray}

\begin{eqnarray}
\Delta^{sv,(0)}_{b,S}&=&\delta(1-z)
\\[2ex]
   \Delta^{sv,(1)}_{b,S} &=&
C_F \Bigg(
        \delta(1-z)      \Bigg[
          - 4
          + 8  \zeta_2
          \Bigg]
       + {\cal D}_1      \Bigg[
           16
          \Bigg]
\Bigg)
\\[2ex]
   \Delta^{sv,(2)}_{b,S} &=&
\delta(1-z) \Bigg[ 
          C_F  C_A    \Bigg(
           {166\over 9}
          - 8  \zeta_3
          + {232\over 9}  \zeta_2
          - {12\over 5}  \zeta_2^2
          \Bigg)
       +   C_F^2    \Bigg(
           16
          - 60  \zeta_3
          + {8\over 5}  \zeta_2^2
          \Bigg)
\nonumber\\[2ex]
&&       +   n_f  C_F    \Bigg(
           {8\over 9}
          + 8  \zeta_3
          - {40\over 9}  \zeta_2
          \Bigg)
\Bigg]
+ {\cal D}_0 \Bigg[ 
           C_F  C_A    \Bigg(
          - {1616\over 27}
          + 56  \zeta_3
          + {176\over 3}  \zeta_2
          \Bigg)
\nonumber\\[2ex]
&&       +   C_F^2    \Bigg(
           256  \zeta_3
          \Bigg)
       +   n_f  C_F    \Bigg(
           {224\over 27}
          - {32\over 3}  \zeta_2
          \Bigg)
\Bigg]
+ {\cal D}_1 \Bigg[ 
          C_F  C_A    \Bigg(
           {1072\over 9}
          - 32  \zeta_2
          \Bigg)
\nonumber\\[2ex]
&&       +   C_F^2    \Bigg(
          - 64
          - 128  \zeta_2
          \Bigg)
       +   n_f  C_F    \Bigg(
          - {160\over 9}
          \Bigg)
\Bigg]
+ {\cal D}_2 \Bigg[  
           C_F  C_A    \Bigg(
          - {176\over 3}
          \Bigg)
       +   n_f  C_F    \Bigg(
           {32\over 3}
          \Bigg)
\Bigg]
\nonumber\\[2ex]
&&+ {\cal D}_3  C_F^2    \Bigg[
           128
          \Bigg]
\\[2ex]
   \Delta^{sv,(3)}_{b,S} &=&
{\cal D}_0  \Bigg[
           C_F  C_A^2    \Bigg(
          - {594058\over 729}
          - 384  \zeta_5
          + {40144\over 27}  \zeta_3
          + {98224\over 81}  \zeta_2
          - {352\over 3}  \zeta_2  \zeta_3
          - {2992\over 15}  \zeta_2^2
          \Bigg)
\nonumber\\[2ex]
&&       +  C_F^2  C_A    \Bigg(
           {6464\over 27}
          + {32288\over 9}  \zeta_3
          + {6592\over 27}  \zeta_2
          - 1472  \zeta_2  \zeta_3
          + {1408\over 3}  \zeta_2^2
          \Bigg)
       +  C_F^3    \Bigg(
           12288  \zeta_5
\nonumber\\[2ex]
&&          - 1024  \zeta_3
          - 6144  \zeta_2  \zeta_3
          \Bigg)
       +  n_f  C_F  C_A    \Bigg(
           {125252\over 729}
          - {2480\over 9}  \zeta_3
          - {29392\over 81}  \zeta_2
          + {736\over 15}  \zeta_2^2
          \Bigg)
\nonumber\\[2ex]
&&       +  n_f  C_F^2    \Bigg(
           {842\over 9}
          - {5728\over 9}  \zeta_3
          - {1504\over 27}  \zeta_2
          - {1472\over 15}  \zeta_2^2
          \Bigg)
       +  n_f^2  C_F    \Bigg(
          - {3712\over 729}
          + {320\over 27}  \zeta_3
\nonumber\\[2ex]
&&          + {640\over 27}  \zeta_2
          \Bigg)
\Bigg]
+ {\cal D}_1 \Bigg[  
            C_F  C_A^2    \Bigg(
           {124024\over 81}
          - 704  \zeta_3
          - {12032\over 9}  \zeta_2
          + {704\over 5}  \zeta_2^2
          \Bigg)
\nonumber\\[2ex]
&&       +   C_F^2  C_A    \Bigg(
          - {544\over 3}
          - 5760  \zeta_3
          - {20864\over 9}  \zeta_2
          + {3648\over 5}  \zeta_2^2
          \Bigg)
       +   C_F^3    \Bigg(
           256
          - 960  \zeta_3
          + 1024  \zeta_2
\nonumber\\[2ex]
&&          - {14208\over 5}  \zeta_2^2
          \Bigg)
       +  n_f  C_F  C_A    \Bigg(
          - {32816\over 81}
          + 384  \zeta_2
          \Bigg)
       +  n_f  C_F^2    \Bigg(
          - {184\over 3}
          + 1280  \zeta_3
          + {3200\over 9}  \zeta_2
          \Bigg)
\nonumber\\[2ex]
&&       +  n_f^2  C_F    \Bigg(
           {1600\over 81}
          - {256\over 9}  \zeta_2
          \Bigg)
\Bigg]
+ {\cal D}_2 \Bigg[
           C_F  C_A^2    \Bigg(
          - {28480\over 27}
          + {704\over 3}  \zeta_2
          \Bigg)
       +   C_F^2  C_A    \Bigg(
          - {10816\over 9}
\nonumber\\[2ex]
&&          + 1344  \zeta_3
          + {11264\over 3}  \zeta_2
          \Bigg)
       +  C_F^3    \Bigg(
           10240  \zeta_3
          \Bigg)
       +  n_f  C_F  C_A    \Bigg(
           {9248\over 27}
          - {128\over 3}  \zeta_2
          \Bigg)
\nonumber\\[2ex]
&&       +  n_f  C_F^2    \Bigg(
           {1696\over 9}
          - {2048\over 3}  \zeta_2
          \Bigg)
       +  n_f^2  C_F    \Bigg(
          - {640\over 27}
          \Bigg)
\Bigg]
+ {\cal D}_3 \Bigg[ 
          C_F  C_A^2    \Bigg(
           {7744\over 27}
          \Bigg)
\nonumber\\[2ex]
&&       +  C_F^2  C_A    \Bigg(
           {17152\over 9}
          - 512  \zeta_2
          \Bigg)
       +  C_F^3    \Bigg(
          - 512
          - 3072  \zeta_2
          \Bigg)
       +  n_f  C_F  C_A    \Bigg(
          - {2816\over 27}
          \Bigg)
\nonumber\\[2ex]
&&       +  n_f  C_F^2    \Bigg(
          - {2560\over 9}
          \Bigg)
       +  n_f^2  C_F    \Bigg(
           {256\over 27}
          \Bigg)
\Bigg]
+ {\cal D}_4 \Bigg[  
            C_F^2  C_A    \Bigg(
          - {7040\over 9}
          \Bigg)
       +  n_f  C_F^2    \Bigg(
           {1280\over 9}
          \Bigg)
\Bigg]
\nonumber\\[2ex]
&&+ {\cal D}_5  C_F^3    \Bigg[
           512
          \Bigg]
\\[2ex]
   \Delta^{sv,(4)}_{b,S} &=&
{\cal D}_2  \Bigg[
            C_F  C_A^3    \Bigg(
          - {1651520\over 81}
          + 7744  \zeta_3
          + {138880\over 9}  \zeta_2
          - {7744\over 5}  \zeta_2^2
          \Bigg)
       +  C_F^2  C_A^2    \Bigg(
          - {6588656\over 243}
\nonumber\\[2ex]
&&          - 9216  \zeta_5
          + {365632\over 3}  \zeta_3
          + {2785408\over 27}  \zeta_2
          - 5504  \zeta_2  \zeta_3
          - {337216\over 15}  \zeta_2^2
          \Bigg)
       +   C_F^3  C_A    \Bigg(
           {43264\over 9}
\nonumber\\[2ex]
&&          + {680512\over 3}  \zeta_3
          + {52736\over 3}  \zeta_2
          - 93696  \zeta_2  \zeta_3
          + {437888\over 15}  \zeta_2^2
          \Bigg)
       +   C_F^4    \Bigg(
           688128  \zeta_5
          - 40960  \zeta_3
\nonumber\\[2ex]
&&          - 409600  \zeta_2  \zeta_3
          \Bigg)
       +   n_f  C_F  C_A^2    \Bigg(
           {82216\over 9}
          - 1408  \zeta_3
          - {64000\over 9}  \zeta_2
          + {1408\over 5}  \zeta_2^2
          \Bigg)
\nonumber\\[2ex]
&&       +   n_f  C_F^2  C_A    \Bigg(
           {2004352\over 243}
          - {338432\over 9}  \zeta_3
          - 32640  \zeta_2
          + {13184\over 3}  \zeta_2^2
          \Bigg)
       +   n_f  C_F^3    \Bigg(
           2272
\nonumber\\[2ex]
&&          - {109184\over 3}  \zeta_3
          - {11264\over 3}  \zeta_2
          - {84224\over 15}  \zeta_2^2
          \Bigg)
       +   n_f^2  C_F  C_A    \Bigg(
          - {31576\over 27}
          + {9728\over 9}  \zeta_2
          \Bigg)
\nonumber\\[2ex]
&&       +   n_f^2  C_F^2    \Bigg(
          - {151424\over 243}
          + {28160\over 9}  \zeta_3
          + {62720\over 27}  \zeta_2
          \Bigg)
       +   n_f^3  C_F    \Bigg(
           {3200\over 81}
          - {512\over 9}  \zeta_2
          \Bigg)
\Bigg]
\nonumber\\[2ex]
&&+ {\cal D}_3 \Bigg[  
            C_F  C_A^3    \Bigg(
           {698368\over 81}
          - {15488\over 9}  \zeta_2
          \Bigg)
       +  C_F^2  C_A^2    \Bigg(
           {8535808\over 243}
          - {140800\over 9}  \zeta_3
          - 52736  \zeta_2
\nonumber\\[2ex]
&&          + {13824\over 5}  \zeta_2^2
          \Bigg)
       +   C_F^3  C_A    \Bigg(
          - {47360\over 9}
          - {1541120\over 9}  \zeta_3
          - {637952\over 9}  \zeta_2
          + {100864\over 5}  \zeta_2^2
          \Bigg)
\nonumber\\[2ex]
&&       +  C_F^4    \Bigg(
           2048
          - 7680  \zeta_3
          + 16384  \zeta_2
          - {195584\over 5}  \zeta_2^2
          \Bigg)
       +  n_f  C_F  C_A^2    \Bigg(
          - {117184\over 27}
          + {5632\over 9}  \zeta_2
          \Bigg)
\nonumber\\[2ex]
&&       +  n_f  C_F^2  C_A    \Bigg(
          - {2447936\over 243}
          + {7168\over 9}  \zeta_3
          + {50176\over 3}  \zeta_2
          \Bigg)
       +  n_f  C_F^3    \Bigg(
          - {9856\over 9}
          + {306176\over 9}  \zeta_3
\nonumber\\[2ex]
&&          + {97280\over 9}  \zeta_2
          \Bigg)
       +  n_f^2  C_F  C_A    \Bigg(
           {18304\over 27}
          - {512\over 9}  \zeta_2
          \Bigg)
       +  n_f^2  C_F^2    \Bigg(
           {151936\over 243}
          - {4096\over 3}  \zeta_2
          \Bigg)
\nonumber\\[2ex]
&&       +  n_f^3  C_F    \Bigg(
          - {2560\over 81}
          \Bigg)
\Bigg]
+ {\cal D}_4 \Bigg[  
           C_F  C_A^3    \Bigg(
          - {42592\over 27}
          \Bigg)
       +  C_F^2  C_A^2    \Bigg(
          - {536960\over 27}
          + {14080\over 3}  \zeta_2
          \Bigg)
\nonumber\\[2ex]
&&       +  C_F^3  C_A    \Bigg(
          - {174080\over 27}
          + 8960  \zeta_3
          + {478720\over 9}  \zeta_2
          \Bigg)
       +  C_F^4    \Bigg(
           {286720\over 3}  \zeta_3
          \Bigg)
\nonumber\\[2ex]
&&       +  n_f  C_F  C_A^2    \Bigg(
           {7744\over 9}
          \Bigg)
       +  n_f  C_F^2  C_A    \Bigg(
           {175360\over 27}
          - {2560\over 3}  \zeta_2
          \Bigg)
       +  n_f  C_F^3    \Bigg(
           {32000\over 27}
\nonumber\\[2ex]
&&          - {87040\over 9}  \zeta_2
          \Bigg)
       +  n_f^2  C_F  C_A    \Bigg(
          - {1408\over 9}
          \Bigg)
       +  n_f^2  C_F^2    \Bigg(
          - {12800\over 27}
          \Bigg)
       +  n_f^3  C_F    \Bigg(
           {256\over 27}
          \Bigg)
\Bigg]
\nonumber\\[2ex]
&&+ {\cal D}_5 \Bigg[  
           C_F^2  C_A^2    \Bigg(
           {123904\over 27}
          \Bigg)
       +  C_F^3  C_A    \Bigg(
           {34304\over 3}
          - 3072  \zeta_2
          \Bigg)
       +  C_F^4    \Bigg(
          - 2048
          - 20480  \zeta_2
          \Bigg)
\nonumber\\[2ex]
&&       +  n_f  C_F^2  C_A    \Bigg(
          - {45056\over 27}
          \Bigg)
       +  n_f  C_F^3    \Bigg(
          - {5120\over 3}
          \Bigg)
       +  n_f^2  C_F^2    \Bigg(
           {4096\over 27}
          \Bigg)
\Bigg]
\nonumber\\[2ex]
&&+ {\cal D}_6 \Bigg[  
           C_F^3  C_A    \Bigg(
          - {39424\over 9}
          \Bigg)
       +  n_f  C_F^3    \Bigg(
           {7168\over 9}
          \Bigg)
\Bigg]
+ {\cal D}_7  C_F^4    \Bigg[
           {4096\over 3}
          \Bigg]
\end{eqnarray}
\begin{eqnarray}
\Delta_{g,S}^{sv,(0)} &=& \delta(1-z)
\\[2ex]
\Delta_{g,S}^{sv,(1)} &=& 
C_A \Bigg(
        \delta(1-z)    \Bigg[
           8  \zeta_2
          \Bigg]
       + {\cal D}_1    \Bigg[
           16
          \Bigg]
\Bigg)
\\[2ex]
\Delta_{g,S}^{sv,(2)} &=& 
\delta(1-z) \Bigg[
         n_f  C_F   \Bigg(
          - {67  \over 3}
          + 16  \zeta_3
          \Bigg)
       +  n_f  C_A   \Bigg(
          - {80  \over 3}
          - {80  \over 9}  \zeta_2
          - {8  \over 3}  \zeta_3
          \Bigg)
       +   C_A^2   \Bigg(
           93
          + {536  \over 9}  \zeta_2
\nonumber\\[2ex]
&&          - {4  \over 5}  \zeta_2^2
          - {220  \over 3}  \zeta_3
          \Bigg)
\Bigg]
+ {\cal D}_0  \Bigg[
       n_f  C_A   \Bigg(
           {224  \over 27}
          - {32  \over 3}  \zeta_2
          \Bigg)
       +   C_A^2   \Bigg(
          - {1616  \over 27}
          + {176  \over 3}  \zeta_2
          + 312  \zeta_3
          \Bigg)
\Bigg]
\nonumber\\[2ex]
&&+ {\cal D}_1 \Bigg[  
          n_f  C_A   \Bigg(
          - {160  \over 9}
          \Bigg)
       +  C_A^2   \Bigg(
           {1072  \over 9}
          - 160  \zeta_2
          \Bigg)
\Bigg]
+ {\cal D}_2 \Bigg[ 
          n_f  C_A   \Bigg(
           {32  \over 3}
          \Bigg)
       +   C_A^2   \Bigg(
          - {176  \over 3}
          \Bigg)
\Bigg]
\nonumber\\[2ex]
&&+ {\cal D}_3  C_A^2   \Bigg[
           128
          \Bigg]
\\[2ex]
\Delta_{g,S}^{sv,(3)} &=&
{\cal D}_0  \Bigg[ 
          n_f  C_F  C_A   \Bigg(
           {3422  \over 27}
          - 32  \zeta_2
          - {64  \over 5}  \zeta_2^2
          - {608  \over 9}  \zeta_3
          \Bigg)
       +   n_f  C_A^2   \Bigg(
           {125252  \over 729}
          - {34768  \over 81}  \zeta_2
\nonumber\\[2ex]
&&          - {544  \over 15}  \zeta_2^2
          - {7600  \over 9}  \zeta_3
          \Bigg)
       +   n_f^2  C_A   \Bigg(
          - {3712  \over 729}
          + {640  \over 27}  \zeta_2
          + {320  \over 27}  \zeta_3
          \Bigg)
       +   C_A^3   \Bigg(
          - {594058  \over 729}
\nonumber\\[2ex]
&&          - {23200  \over 3}  \zeta_2  \zeta_3
          + {137008  \over 81}  \zeta_2
          + {4048  \over 15}  \zeta_2^2
          + {143056  \over 27}  \zeta_3
          + 11904  \zeta_5
          \Bigg)
\Bigg]
\nonumber\\[2ex]
&&+ {\cal D}_1  \Bigg[
          n_f  C_F  C_A   \Bigg(
          - 504
          + 384  \zeta_3
          \Bigg)
       +   n_f  C_A^2   \Bigg(
          - {67376  \over 81}
          + {6016  \over 9}  \zeta_2
          + {2944  \over 3}  \zeta_3
          \Bigg)
\nonumber\\[2ex]
&&       +  n_f^2  C_A   \Bigg(
           {1600  \over 81}
          - {256  \over 9}  \zeta_2
          \Bigg)
       +  C_A^3   \Bigg(
           {244552  \over 81}
          - {9728  \over 3}  \zeta_2
          - {9856  \over 5}  \zeta_2^2
          - {22528  \over 3}  \zeta_3
          \Bigg)
\Bigg]
\nonumber\\[2ex]
&&+ {\cal D}_2 \Bigg[ 
          n_f  C_F  C_A   \Bigg(
           32
          \Bigg)
       +   n_f  C_A^2   \Bigg(
           {14624  \over 27}
          - {2176  \over 3}  \zeta_2
          \Bigg)
       +  n_f^2  C_A   \Bigg(
          - {640  \over 27}
          \Bigg)
\nonumber\\[2ex]
&&       +  C_A^3   \Bigg(
          - {67264  \over 27}
          + {11968  \over 3}  \zeta_2
          + 11584  \zeta_3
          \Bigg)
\Bigg]
+ {\cal D}_3  \Bigg[
         n_f  C_A^2   \Bigg(
          - {10496  \over 27}
          \Bigg)
\nonumber\\[2ex]
&&       +   n_f^2  C_A   \Bigg(
           {256  \over 27}
          \Bigg)
       +   C_A^3   \Bigg(
           {59200  \over 27}
          - 3584  \zeta_2
          \Bigg)
\Bigg]
+ {\cal D}_4  \Bigg[
         n_f  C_A^2   \Bigg(
           {1280  \over 9}
          \Bigg)
\nonumber\\[2ex]
&&       +   C_A^3   \Bigg(
          - {7040  \over 9}
          \Bigg)
\Bigg]
+ {\cal D}_5  C_A^3   \Bigg[
           512
          \Bigg]
\\[2ex]
\Delta_{g,S}^{sv,(4)} &=&
 {\cal D}_2 \Bigg[ 
            n_f  C_F  C_A^2   \Bigg(
           6624
          - 2176  \zeta_2
          - {1536  \over 5}  \zeta_2^2
          - 3968  \zeta_3
          \Bigg)
       +   n_f  C_F^2  C_A   \Bigg(
          - 16
          \Bigg)
\nonumber\\[2ex]
&&       +   n_f  C_A^3   \Bigg(
           {4591096  \over 243}
          - {1186688  \over 27}  \zeta_2
          - {9472  \over 15}  \zeta_2^2
          - 71808  \zeta_3
          \Bigg)
       +   n_f^2  C_F  C_A   \Bigg(
          - {5600  \over 9}
\nonumber\\[2ex]
&&          + {1280  \over 3}  \zeta_3
          \Bigg)
       +   n_f^2  C_A^2   \Bigg(
          - {436760  \over 243}
          + {30208  \over 9}  \zeta_2
          + {24832  \over 9}  \zeta_3
          \Bigg)
       +   n_f^3  C_A   \Bigg(
           {3200  \over 81}
\nonumber\\[2ex]
&&          - {512  \over 9}  \zeta_2
          \Bigg)
       +   C_A^4   \Bigg(
          - {13631360  \over 243}
          - 508800  \zeta_2  \zeta_3
          + {4104704  \over 27}  \zeta_2
          + {15488  \over 3}  \zeta_2^2
\nonumber\\[2ex]
&&          + {3259328  \over 9}  \zeta_3
          + 678912  \zeta_5
          \Bigg)
\Bigg]
+ {\cal D}_3  \Bigg[
          n_f  C_F  C_A^2   \Bigg(
          - {50368  \over 9}
          + 4096  \zeta_3
          \Bigg)
\nonumber\\[2ex]
&&       +  n_f  C_A^3   \Bigg(
          - {4338368  \over 243}
          + {248320  \over 9}  \zeta_2
          + {94208  \over 3}  \zeta_3
          \Bigg)
       +   n_f^2  C_F  C_A   \Bigg(
           {640  \over 9}
          \Bigg)
\nonumber\\[2ex]
&&       +   n_f^2  C_A^2   \Bigg(
           {308608  \over 243}
          - {12800  \over 9}  \zeta_2
          \Bigg)
       +   n_f^3  C_A   \Bigg(
          - {2560  \over 81}
          \Bigg)
       +   C_A^4   \Bigg(
           {13802368  \over 243}
          - {1107584  \over 9}  \zeta_2
\nonumber\\[2ex]
&&          - {80896  \over 5}  \zeta_2^2
          - {585728  \over 3}  \zeta_3
          \Bigg)
\Bigg]
+ {\cal D}_4 \Bigg[ 
          n_f  C_F  C_A^2   \Bigg(
           {1280  \over 3}
          \Bigg)
       +   n_f  C_A^3   \Bigg(
           {26048  \over 3}
          - {94720  \over 9}  \zeta_2
          \Bigg)
\nonumber\\[2ex]
&&       +   n_f^2  C_A^2   \Bigg(
          - {17024  \over 27}
          \Bigg)
       +   n_f^3  C_A   \Bigg(
           {256  \over 27}
          \Bigg)
       +   C_A^4   \Bigg(
          - {838112  \over 27}
          + {520960  \over 9}  \zeta_2
          + {313600  \over 3}  \zeta_3
          \Bigg)
\Bigg]
\nonumber\\[2ex]
&&+ {\cal D}_5 \Bigg[ 
          n_f  C_A^3   \Bigg(
          - {91136  \over 27}
          \Bigg)
       +   n_f^2  C_A^2   \Bigg(
           {4096  \over 27}
          \Bigg)
       +   C_A^4   \Bigg(
           {432640  \over 27}
          - 23552  \zeta_2
          \Bigg)
\Bigg]
\nonumber\\[2ex]
&&+ {\cal D}_6  \Bigg[
          n_f  C_A^3   \Bigg(
           {7168  \over 9}
          \Bigg)
       +   C_A^4   \Bigg(
          - {39424  \over 9}
          \Bigg)
\Bigg]
+ {\cal D}_7  C_A^4   \Bigg[
           {4096  \over 3}
          \Bigg]
\end{eqnarray}


\end{document}